\documentclass[times,twocolumn,final]{elsarticle}

\usepackage{medima}
\usepackage{framed,multirow}

\usepackage{amssymb}
\usepackage{latexsym}
\usepackage{url}
\usepackage{amsmath}
\usepackage{todonotes}
\usepackage{algorithmic}
 \usepackage{multirow}
\usepackage{graphicx}
\usepackage{booktabs}
\usepackage{enumitem}
\usepackage{float}
\usepackage{soul}
\usepackage{mathtools}
\usepackage{amsmath}
\usepackage{amsfonts}
\usepackage{bm}
\usepackage{amssymb}
\usepackage{stackengine}
\usepackage{todonotes}
\usepackage{balance}

\usepackage{url}
\usepackage{xcolor}
\DeclareMathOperator{\E}{\mathbb{E}}

\usepackage{hyperref}

\definecolor{newcolor}{rgb}{.8,.349,.1}
\tikzset{/tikz/notestyleraw/.append style={text=black!70!white}}
\journal{Arxiv}

\begin{document}

\verso{Tian Xia, \textit{et~al.}}

\begin{frontmatter}

\title{Learning to synthesise the ageing brain without longitudinal data\tnoteref{tnote1}}%

\author[1]{Tian \snm{Xia}\corref{cor1}}
\cortext[cor1]{Corresponding author. }
\ead{tian.xia@ed.ac.uk}
\author[1]{Agisilaos \snm{Chartsias}}
\author[2]{Chengjia \snm{Wang}}
\author[1,3]{Sotirios A. \snm{Tsaftaris}}
\author[]{for the Alzheimer's Disease Neuroimaging Initiative}

\address[1]{Institute for Digital Communications, School of Engineering, University of Edinburgh, West Mains Rd, Edinburgh EH9 3FB, UK}

\address[2]{The BHF Centre for Cardiovascular Science, Edinburgh EH16 4TJ, UK}

\address[3]{The Alan Turing Institute, London NW1 2DB, UK}

\begin{abstract}
How will my face look when I get older?  Or, for a more challenging question: How will my brain look when I get older? To answer this question one must devise (and learn from data) a multivariate auto-regressive function which given an image and a desired target age generates an output image. While collecting data for faces may be easier, collecting longitudinal brain data is not trivial.  
We propose a deep learning-based method that learns to simulate subject-specific brain ageing trajectories \textit{without} relying on longitudinal data. Our method synthesises images conditioned on two factors: age (a continuous variable), and status of Alzheimer's Disease (AD, an ordinal variable). With an adversarial formulation we learn the joint distribution of brain appearance, age and AD status, and define reconstruction losses to address the challenging problem of preserving subject identity. We compare with several benchmarks using two widely used datasets. We evaluate the quality and realism of synthesised images using ground-truth longitudinal data and a pre-trained age predictor. We show that, despite the use of cross-sectional data, our model learns patterns of gray matter atrophy in the middle temporal gyrus in patients with AD. To demonstrate generalisation ability, we train on one dataset and evaluate predictions on the other. In conclusion, our model shows an ability to separate age, disease influence and anatomy using only 2D cross-sectional data that should be useful in large studies into neurodegenerative disease, that aim to combine several data sources. To facilitate such future studies by the community at large our code is made available at \url{https://github.com/xiat0616/BrainAgeing}. This preprint has been published in Medical Image Analysis: \url{https://doi.org/10.1016/j.media.2021.102169}.

\end{abstract}

\begin{keyword}

\KWD Brain Ageing\sep Generative Adversarial Network\sep Neurodegenerative Disease\sep Magnetic Resonance Imaging (MRI)

\end{keyword}

\end{frontmatter}

\section{Introduction}

\begin{figure*}[t!]

    \centering
    \includegraphics[scale=0.55]{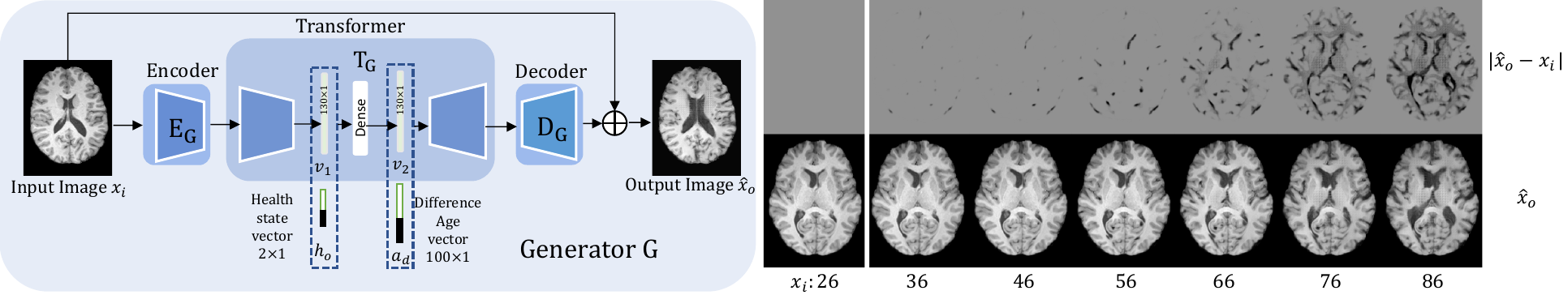}
    \caption{\textbf{Left:} The input is a brain image $x_{i}$, and the network synthesises an aged brain image $\hat{x}_{o}$ from $x_{i}$, conditioned on the target health state vector $h_{o}$ and target age difference $a_{d}=a_{o}-a_{i}$ between input $a_{i}$ and target $a_{o}$ ages, respectively. \textbf{Right: }  For an image $x_i$ of a 26 year old subject, bottom row shows outputs $\hat{x}_{o}$ given different target age. The top row shows the corresponding image differences $|\hat{x}_{o}-x_{i}|$ to highlight progressive changes. }

    \label{fig: simple illustration}
\end{figure*}

The ability to predict the future state of an individual can be of great benefit for longitudinal studies~\citep{ziegler2012models}. However, such learned phenomenological predictive models need to capture anatomical and physiological changes due to ageing and separate the factors that influence future state.
Recently, deep generative models have been used to simulate and predict future degeneration of a human brain using existing scans~\citep{ravi2019degenerative,rachmadi2019predicting,rachmadi2020automatic}. However, current methods require considerable amount of longitudinal data to sufficiently approximate an auto-regressive model. Here, we propose a new conditional adversarial training procedure that does \textit{not} require longitudinal data to train. Our approach (shown in Fig.~\ref{fig: simple illustration}) synthesises images of aged brains for a desired age and health state. 

Brain ageing, accompanied by a series of functional and physiological changes, has been intensively investigated~\citep{zecca2004iron,mattson2018hallmarks}. However, the underlying mechanism has not been completely revealed~\citep{lopez2013hallmarks,cole2019brain}. Prior studies have shown that a brain's chronic changes are related to different factors, e.g.\ the biological age~\citep{fjell2010structural}, degenerative diseases such as Alzheimer's Disease (AD)~\citep{jack1998rate}, binge drinking~\citep{coleman2014adolescent}, and even education~\citep{taubert2010dynamic}. Accurate simulation of this process has great value for both neuroscience research and clinical applications to identify age-related pathologies \citep{cole2019brain,fjell2010structural}. 

One particular challenge is inter-subject variation: every individual has a unique ageing trajectory. Previous approaches built a spatio-temporal atlas to predict average brain images at different ages~\citep{davis2010population,huizinga2018spatio}. However, individuals with different health status follow different ageing trajectories. An atlas may not preserve subject-specific characteristics; thus, may preclude accurate modelling of individual trajectories and further investigation on the effect of different factors, e.g.\ age, gender, education, etc~\citep{ravi2019degenerative1}. 
Recent studies proposed subject-specific ageing progression with neural networks~\citep{ravi2019degenerative,rachmadi2019predicting}, although they require longitudinal data to train. Ideally, the longitudinal data should cover a long time span with frequent sampling to ensure stable training.  However, such data are difficult and expensive to acquire, particularly for longer time spans. Even in ADNI~\citep{petersen2010alzheimer}, one of the most well-known large-scale datasets, longitudinal images are acquired at few time points and cover only a few years. Longitudinal data of sufficient time span remain an open challenge.

\footnotetext[1]{A classical computer vision example is generating a human face resembling another individual instead of the input subject. Even with faces, humans find it difficult to assess identity loss. It remains hard to define detailed structural changes during ageing, e.g.\ balding, nose shape change, eye colour change. There are some common patterns that we can expect, such as wrinkles and gray/white hair, but it is difficult to define other more detailed changes. Therefore, even in face ageing, `subject identity' is defined as young and old images should be from the same person. In brain synthesis, it is even more difficult to define `subject identity', as human eyes are less able to visually ascertain brain image identity particularly as modulated by age and pathology. In this paper, we followed a similar analogue of `identity’: a ``synthetic image should be from the same subject as the input image''.}

 In this paper, we build the foundations of a model that can be trained without longitudinal data. A simplified schematic of our model is shown in Fig.~\ref{fig: simple illustration} along with example results. Given a brain image, our model produces a brain of the same subject at target age. The input image is first encoded into a latent space, which is modulated by two vectors representing target age difference and health state (AD status in this paper), respectively. The conditioned latent space is finally decoded to an output image, i.e.\ the synthetically aged image.

 Under the hood, what trains the generator, is a deep adversarial method that learns the joint distribution of brain appearance, age and health state. The quality of the output is encouraged by a discriminator that judges whether an output image is representative of the distribution of brain images of the desired age and health state. A typical problem in synthesis which is exacerbated with \textit{cross-sectional} data \citep{ziegler2012models} is loss of \textit{subject identity}\footnotemark, i.e.\ the synthesis of an output that may not correspond to the input subject's identity. We propose, and motivate, two loss functions towards retaining \textit{subject identity} by regularising the amount of change introduced by ageing.  In addition, we motivate the design of our conditioning mechanisms and show that ordinal binary encoding for both discrete and continuous variables improves performance significantly.

We consider several metrics and evaluation approaches to verify the quality and biological plausibility of our results.  We quantitatively evaluate our simulation results using longitudinal data from the ADNI dataset~\citep{petersen2010alzheimer} with classical metrics that estimate image fidelity. Since the longitudinal data only cover a limited time span, it is difficult to evaluate the quality of synthesized aged images across decades. For brain ageing synthesis, a good synthetic brain image should be accurate in terms of age, i.e.\ be close to the target age that we want it to be, and also preserve subject identity, i.e.\ should be from the same subject as the input. Thus, we pre-train a deep network to estimate the apparent age from output images. The estimated ages are used as a proxy metric for ~\textit{age accuracy}. We also show qualitative results, including  ageing simulation on different health states and long-term ageing synthesis. Both quantitative and qualitative results show that our method outperforms benchmarks with more accurate simulations that capture the characteristics specific to each individual on different health states. Furthermore, we train our model on Cam-CAN and evaluate it on ADNI to demonstrate the generalisation ability  to unseen data. In addition, to demonstrate the realism of synthetic results, we perform volume synthesis and evaluate deformation. We also estimate gray matter atrophy in middle temporal gyrus and find that our model, even without longitudinal data, has learned that ageing and disease leads to atrophy.
Ablation studies investigate the effect of loss components and different ways of embedding clinical variables into the networks. 

Our contributions are summarised as follows:\footnote{We advance our preliminary work~\citep{xia2019consistent} in the following aspects:  1) we extend our model to condition on age and AD status, which enables more accurate simulation of ageing progression of different health states; 2) we introduce additional regularisation to smooth the simulated progression; 3) we offer more experiments and a detailed analysis of performance, using longitudinal data, including new metrics and additional benchmark methods for comparison; 4) we introduce analysis based on measuring deformation and atrophy; and 5) several ablation studies. }
\begin{itemize}
    \item Our main contribution is a deep learning model that learns to simulate the brain ageing process, and perform subject-specific brain ageing synthesis, trained on \textit{cross-sectional} data overcoming the need for longitudinal data. 
    \item For our model to be able to change output based on desired input (age and health state), we use an (ordinal) embedding mechanism that guides the network to learn the joint distribution of brain images, age and health state.
    \item Since we do not use longitudinal data that can constrain the learning process, we design losses that aim to preserve subject identity, while encouraging quality output. 
    \item We provide an experimental framework to verify the quality and biological validity of the  synthetic outputs. 
\end{itemize}
While our first contribution is the most important one, it is the combination of our proposed losses and embedding mechanisms that lead to the method's robustness, as extensive experiments and ablation studies on two publicly available datasets, namely Cam-CAN~\citep{camcan2015ageing}  and ADNI~\citep{petersen2010alzheimer} show.

The manuscript proceeds as follows: Section~\ref{sec: related work} reviews related work on brain ageing simulation and prediction.  
Section~\ref{sec: proposed approach} details the proposed method. Section~\ref{sec: experimenral setup} describes the experimental setup and training details. Section~\ref{sec: results and discussion} presents results and discussion. Finally, Section~\ref{sec: conclusion} offers conclusions.

\section{Related Work}
\label{sec: related work}
 We first discuss \textit{brain ageing simulation}, i.e.\ simulating  the ageing process from data. For completeness, we also briefly discuss \textit{brain age prediction}, i.e.\ estimating age from an image.

\subsection{Brain ageing simulation}  Given variables such as age, one can synthesise the corresponding brain image to enable visual observation of brain changes. For instance, patch-based dictionary learning~\citep{zhang2016consistent}, kernel regression~\citep{huizinga2018spatio,ziegler2012models, serag2012construction}, linear mixed-effect modelling~\citep{lorenzi2015disentangling,sivera2019model} and non-rigid registration~\citep{sharma2010evaluation,modat2014simulating,pieperhoff2008detection,camara2006phenomenological}  have been used to build spatio-temporal atlases of brains at different ages. However, by relying on population averages as atlases subject-specific ageing trajectories are harder to capture. Recently,~\citet{khanal2017simulating} build a biophysical model assuming brain atrophy, but without considering age or other clinical factors (e.g.\ AD status).


Deep generative methods have also been used for this task. While \citet{rachmadi2019predicting,rachmadi2020automatic} and \citet{wegmayr2019generative} used formulations of Generative Adversarial Networks (GAN)~\citep{goodfellow2014generative} to simulate brain changes, others \citep{ravi2019degenerative} used a conditional adversarial autoencoder as the generative model, following a recent face ageing approach~\citep{zhang2017age}.  Irrespective of the model, these methods need longitudinal data, which limits their applicability. 

In~\citet{bowles2018modelling}, a GAN-based method is trained to add or remove atrophy patterns in the brain using image arithmetics. However, the atrophy patterns were  modelled linearly, with  morphological changes assumed to be the same for all subjects.
 \cite{milana2017deep} used a Variational Autoencoder (VAE) to synthesise aged brain images, but the target age is not controlled, and the quality of the synthesised image appears poor (blurry). Recently, \cite{pawlowski2020deep} showed that a VAE-based structural causal model can generate brain images. However, they did not provide a quantitative evaluation of the generated images perhaps due to known issues of low quality outputs when using a VAE. Similarly,~\cite{zhao2019variational} used a VAE to disentangle the spatial information from temporal progression and then used the first few layers of the VAE as feature extractor to improve the age prediction task. As their focus is on age prediction, the synthetic brain images only contain the ventricular region and are population averages.
 
 
 
 In summary, most previous methods either built average atlases~\citep{zhang2016consistent,huizinga2018spatio,ziegler2012models,serag2012construction}, or required longitudinal data~\citep{rachmadi2019predicting,rachmadi2020automatic,ravi2019degenerative,wegmayr2019generative} to simulate brain ageing. Other methods either did not consider subject identity~\citep{bowles2018modelling,milana2017deep}, or did not evaluate in detail morphological changes~\citep{pawlowski2020deep,zhao2019variational}. 
 
To address these shortcomings, we propose a conditional adversarial training procedure that learns to simulate the brain ageing process by being \textit{specific} to the input subject, and by learning from \textit{cross-sectional} data i.e.\ without requiring longitudinal observations. 

\subsection{Brain age prediction} These methods predict age from brain images learning a relationship between image and age; thus, for completeness we briefly mention two key directions. For example,~\citet{franke2010estimating} predicted age with hand-crafted features and kernel regression whereas~\citet{cole2017predicting} used Gaussian Processes. Naturally performance relies on the effectiveness of the hand-crafted features. 

\begin{figure*}[t!]
    \centering
    \includegraphics[scale=0.55]{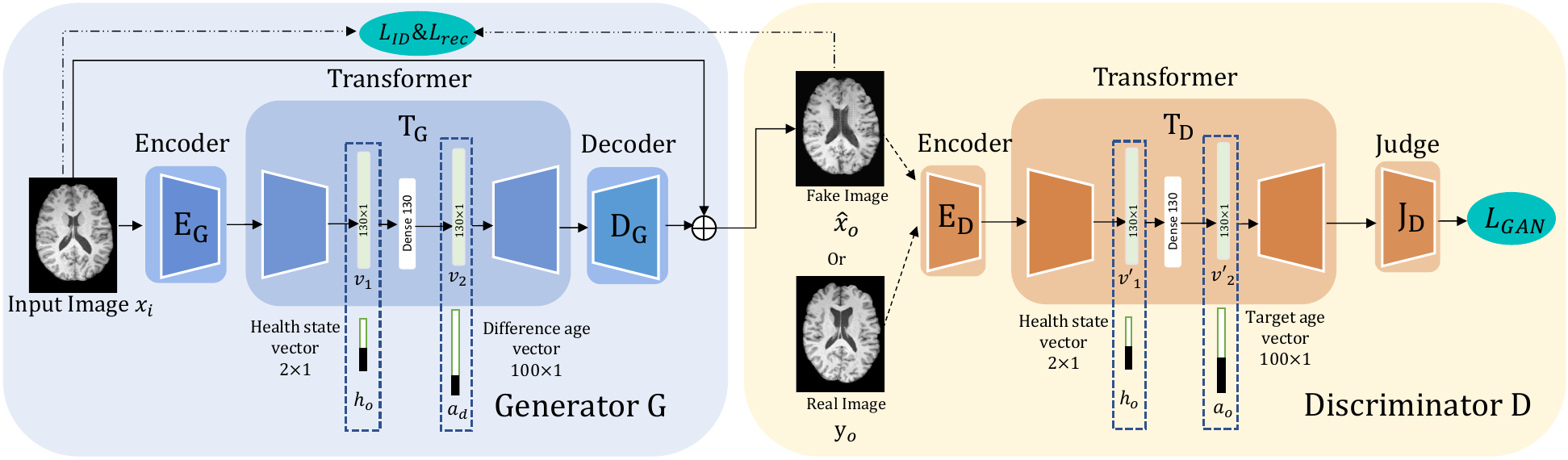}
    \caption{An overview of the proposed method (training). $\mathbf{x}_{i}$ is the input image; $\mathbf{h}_{o}$ is the target health state;  $\mathbf{a}_{d}$ is the difference between the starting age $a_{i}$ and target age $a_{o}$: $a_{d}=a_{o}-a_{i}$;  $\mathbf{\hat{x}}_{o}$ is the output (aged) image (supposedly belong to the same subject as $x_{i}$) of the target age $a_{o}$ and health state $h_{o}$. The \textit{Generator} takes as input $\mathbf{x}_{i}$, $\mathbf{h}_{o}$ and $\mathbf{a}_{d}$, and outputs $\mathbf{\hat{x}}_{o}$; the \textit{Discriminator} takes as input a brain image and  $\mathbf{h}_{o}$ and $\mathbf{a}_{o}$, and outputs a discrimination score.}
    \label{fig: overview of method}
\end{figure*}

Recently, deep learning models have been used to estimate the brain age from imaging data. For example,~\citet{cole2017mortality} used a VGG-based model~\citep{simonyan2014very} to predict age and detect degenerative diseases, while~\citet{jonsson2019deep} proposed to discover genetic associations with the brain degeneration using a ResNet-based network~\citep{he2016deep}. Similarly,~\citet{peng2021accurate} used a CNN-based model to predict age. ~\citet{cole2015prediction} used the age predicted by a deep network to detect traumatic brain injury. While most previous works achieved mean absolute error (MAE) of 4-5 years,~\citet{peng2021accurate} achieved state-of-the-art performance with MAE of 2.14 years. However, these methods did not consider the morphological changes of brain, which is potentially more informative~\citep{costafreda2011automated}.

\section{Proposed approach}
\label{sec: proposed approach}

\subsection{Problem statement, notation and overview}

In the rest of the paper, we use \textbf{bold} notations for vectors/images, and \textit{italics} notations for scalars. For instance, $a$ represents an age while $\mathbf{a}$ is a vector that represents age $a$. We denote a brain image as $\mathbf{x}_{s}$ (and $\mathcal{X}_{s}$ their distribution such that $\mathbf{x}_{s}\sim \mathcal{X}_{s}$), where $s$ are the subject's clinical variables including the corresponding age $a$ and health state (AD status) $h$.  Given a brain image $\mathbf{x}_{i}$ of age $a_{i}$ and health state $h_{i}$, we want to synthesise a brain image $\mathbf{\hat{x}}_{o}$ of target age $a_{o}$ and health state $h_{o}$.  Critically, the synthetic brain image $\mathbf{\hat{x}}_{o}$ should retain the subject identity, i.e.\ belong to the same subject as the input $\mathbf{x}_{i}$, throughout the ageing process. The contributions of our approach, shown in Fig.~\ref{fig: overview of method}, are the design of the conditioning mechanism; our model architecture that uses a Generator to synthesise images, and a Discriminator to help learn the joint distribution of clinical variables and brain appearance; and the losses we use to guide the training process.  We detail all these below.

\subsection{Conditioning on age and health state}
\label{sec: conditioning on age and AD}

In our previous work \citep{xia2019consistent}, we simulate the ageing brain with age as the single factor.  
Here, we improve our previous approach by involving the health state, i.e.\ AD status, as another factor to better simulate the ageing process.
\footnote{Additional fine-grained information on AD effects on different, local, brain regions could be provided if clinical scores are used instead. As our work is the first to attempt to learn without longitudinal data, for simplicity we focused on variables capturing global effects. In the conclusion section, we note the addition of fine-grained information as an avenue for future improvement.}

\begin{figure}[t]
    \centering
    \includegraphics[scale=0.5]{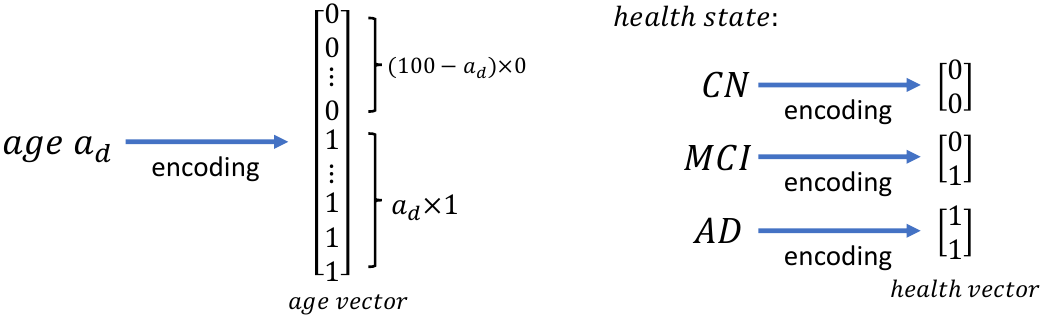}
    \caption{Ordinal encoding of age and health state. \textbf{Left} shows how we represent age $\mathbf{a}_{d}$ using a binary vector with first $\mathbf{a}_{d}$ elements as 1 and the rest as 0; \textbf{Right} is the encoding of health state, where we use a $2\times1$ vector to represent three categories of AD status: control normal (CN), mildly cognitive impaired (MCI), and  Alzheimer's Disease (AD).  }
    \label{fig: ordinal encoding}
\end{figure}

We use ordinal binary vectors, instead of one-hot vectors as in~\citet{zhang2017age}, to encode both age and health state, which are embedded in the bottleneck layer of the Generator and Discriminator (detailed in Section~\ref{sec: model archiecture}). We assume a maximal age of 100 years  and use a $100\times1$ vector to encode age $a$. Similarly, we use a $2\times1$ vector to encode health state. A simple illustration of this encoding is shown in Fig.~\ref{fig: ordinal encoding}. An ablation study presented in Section~\ref{sec: ablation studies} illustrates the benefits of \textit{ordinal} v.s. \textit{one-hot} encoding.

\subsection{Proposed model}
\label{sec: model archiecture}
\begin{figure*}[t!]
    \centering
    \includegraphics[scale=0.55]{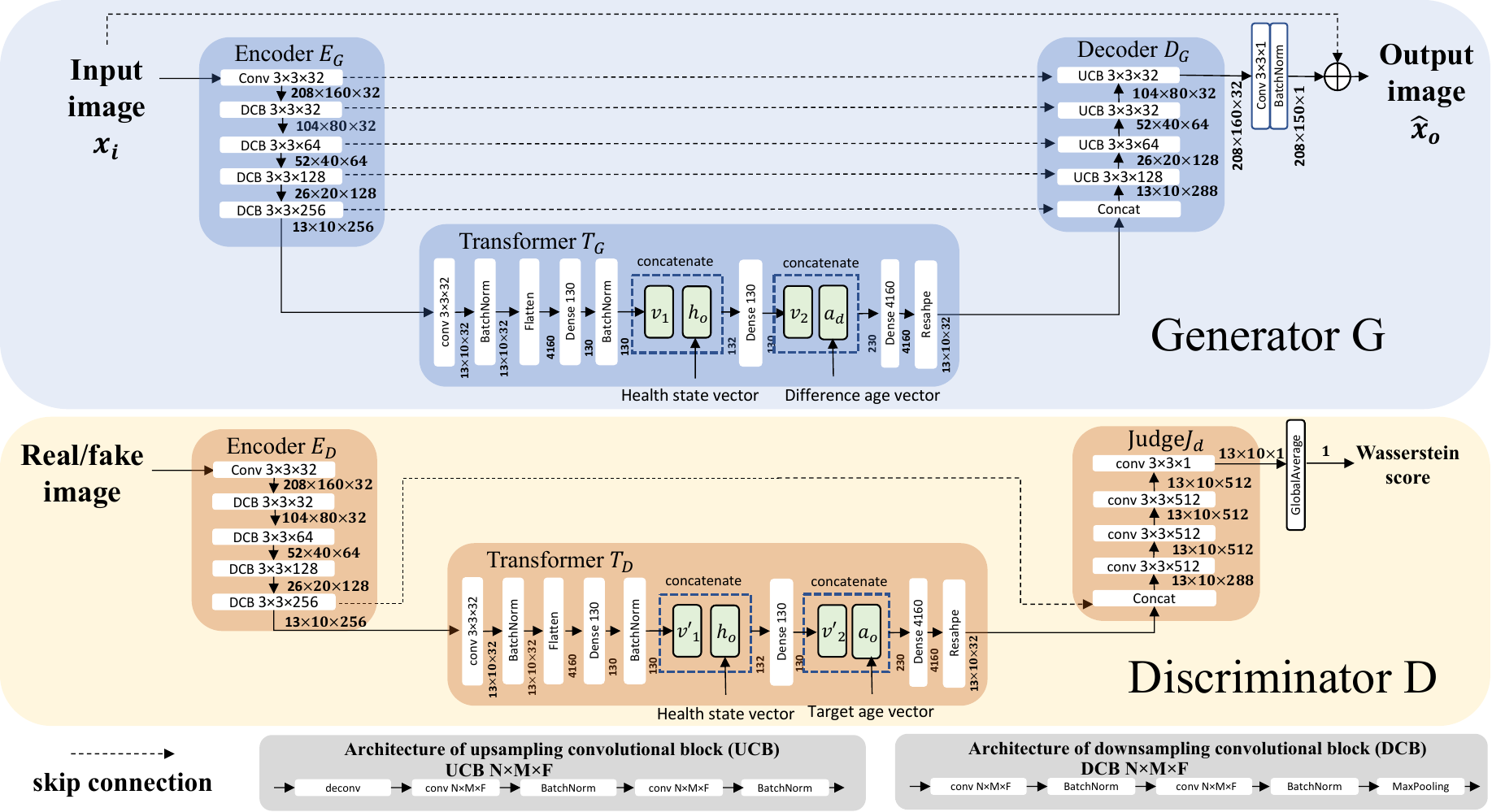}
    \caption{Detailed architectures of \textit{Generator} and \textit{Discriminator}. The Generator contains three parts: an Encoder to extract latent features; a Transformer to involve target age and health state; and a Decoder to generate  aged images. Similarly, we use the same conditioning mechanism for the Discriminator to inject the information of age and health state, and a long skip connection to better preserve features of input image. }
    \label{fig: detailed structures}
\end{figure*}

The proposed method consists of a \textit{Generator} and a \textit{Discriminator}. The Generator synthesises aged brain images corresponding to a target age and a health state.  The Discriminator has a dual role: firstly, it discriminates between ground-truth and synthetic brain images; secondly, it ensures that the synthetic brain images correspond to the target clinical variables. The Generator is adversarially trained to generate realistic brain images of the correct target age. The detailed network architectures are shown in Fig.~\ref{fig: detailed structures}.

\subsubsection{Generator}
The Generator $G$ takes as input a 2D brain image $\mathbf{x}_{i}$, and ordinal binary vectors for target health state $h_{o}$  and age difference $a_{d}$. Here, we condition on the age difference between input age $a_{i}$ and target age $a_{o}$: $a_{d}=a_{o}- a_{i}$, such that when input and output ages are equal $a_{d}=0$, the network is encouraged to recreate the input. The output of $G$ is a 2D brain image $\mathbf{\hat{x}}_{o}$ corresponding to the target age and health state.\footnote{Note that the target health state can be different from the corresponding input state. This  encourages learning a joint distribution between brain images and clinical variables.}

$G$ has three components: the \textit{Encoder} $E_{G}$, the \textit{Transformer} $T_{G}$  and the \textit{Decoder} $D_{G}$. $E_{G}$ first extracts latent features from the input image $\mathbf{x}_{i}$;  $T_{G}$ involves the target age and health state into the network. Finally, $D_{G}$ generates the aged brain image from the bottleneck features. To embed age and health state into our model, we first concatenate the latent vector $\mathbf{v}_{1}$, obtained  by $E_{G}$, with the health state vector $h_{o}$.  The concatenated vector is then processed by a dense layer to output latent vector $\mathbf{v}_{2}$, which is then concatenated with the difference age vector $\mathbf{a}_{d}$. Finally, the resulting vector is used to generate the output image.\footnote{We tested the ordering of $\mathbf{h}_{o}$ and $\mathbf{a}_{d}$, and it did not affect the results. We also tried to concatenate $\mathbf{h}_{o}$, $\mathbf{a}_{d}$ and $\mathbf{v}_{1}$ together into one  vector, and use the resulting vector to generate the output. However, we found that the model tended to ignore the information of $\mathbf{h}_{o}$. This might be caused by the dimensional imbalance between $h_{o}$ ($2\times1$) and $a_{d}$ ($100\times1$).} We adopt long-skip connections~\citep{ronneberger2015u} between layers of $E_{G}$ and $D_{G}$ to preserve details of the input image and improve the sharpness of the output images. Overall, the Generator's forward pass is: $\mathbf{\hat{x}}_{o}=G(\mathbf{x}_{i}, \mathbf{a}_{d}, \mathbf{h}_{o})$.

\subsubsection{Discriminator} Similar to the Generator, the Discriminator $D$ contains three subnetworks: the \textit{Encoder} $E_{D}$ that extracts latent features, the \textit{Transformer} $T_{D}$ that involves the conditional variables, and the \textit{Judge} $J_{D}$ that outputs a discrimination score. For the discriminator to learn the joint distribution  of brain image, age, and health state, we embed the age and health vectors into the discriminator with a similar mechanism as that of the Generator.

Note that  $D$ is conditioned on the target age $\mathbf{a}_{o}$ instead of age difference $\mathbf{a}_{d}$, to learn the joint distribution of brain appearance and age, such that it can discriminate between real and synthetic images of correct age. The forward pass for the Discriminator is $w_{fake}=D(\mathbf{\hat{x}}_{o},\mathbf{a}_{o}, \mathbf{h}_{o})$ and $w_{real}=D(\mathbf{y}_{o},\mathbf{a}_{o}, \mathbf{h}_{o})$.

\subsection{Losses}
We train with a multi-component loss function containing \textit{adversarial}, \textit{identity-preservation} and \textit{self-reconstruction} losses. We detail  these below.

\subsubsection{Adversarial loss}

We adopt the Wasserstein loss with gradient penalty~\citep{gulrajani2017improved} to predict a realistic aged brain image $\mathbf{\hat{x}}_{o}$ and force $\mathbf{\hat{x}}_{o}$ to correspond to the target age $a_{o}$ and health state $h_{o}$:
\begin{equation}
\label{eq: Lgan_1}
\begin{split}
    L_{GAN_{}}=\E_{\mathbf{y}_{o}\sim \mathcal{X}_{o}, \hat{\mathbf{x}}_{o} \sim \hat{\mathcal{X}}_{o}}[D(\mathbf{y}_{o}, \mathbf{a}_{o}, \mathbf{h}_{o} )\\\-D(\hat{\mathbf{x}}_{o}, \mathbf{a}_{o}, \mathbf{h}_{o})+
    \lambda_{GP}({\lVert}\nabla_{\bar{z}} D(\tilde{\mathbf{z}}, \mathbf{a}_{o}, \mathbf{h}_{o}){\rVert}_{2}-1 )_{2}],
\end{split}
\end{equation}
where $\mathbf{\hat{x}}_{o}$ is the output image: $\mathbf{\hat{x}}_{o}=G(\mathbf{x}_{i}, \mathbf{a}_{d}, \mathbf{h}_{o})$ (and $\mathbf{a}_{d}=\mathbf{a}_{o}-\mathbf{a}_{i}$); $\mathbf{y}_{o}$ is a ground truth image from another subject of target age $a_{o}$ and health state ${h_{o}}$; and $\tilde{\mathbf{z}}$ is the average sample defined by $\tilde{\mathbf{z}}=\epsilon \hat{\mathbf{x}}_{o}+(1-\epsilon) \mathbf{y}_{o}, \epsilon \sim U[0,1]$. The first two terms measure the Wasserstein distance between ground-truth and synthetic samples; the last term is the gradient penalty involved to stabilise training. As in~\citet{gulrajani2017improved} and \citet{baumgartner2018visual} we set $\lambda_{GP}=10$.

\begin{figure}[t]
    \centering
    \includegraphics[scale=0.35]{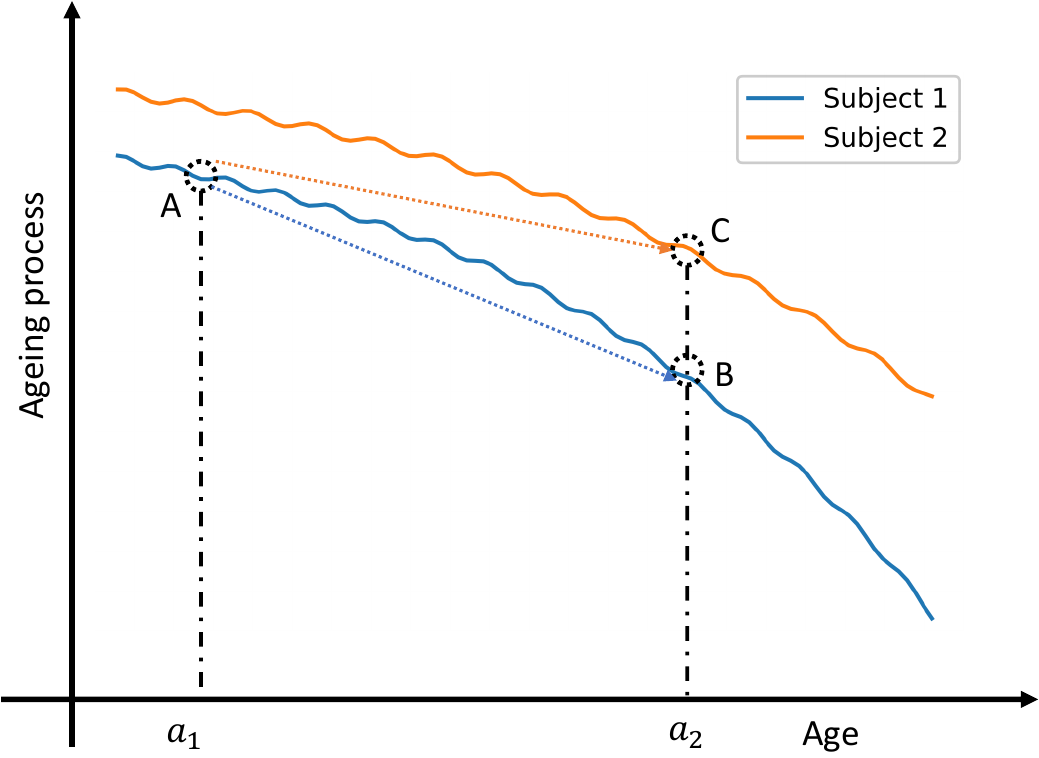}
    \caption{Illustration of ageing trajectories for two subjects. For a subject of age $a_1$ (A), the network can learn a mapping from A to C, which could still fool the Discriminator, but loses the identity of Subject 1 (orange line).}
    \label{fig: illustration of identity loss}
\end{figure}
\subsubsection{Identity-preservation loss}
While $L_{GAN_{}}$  encourages the network to synthesise realistic brain images, these  images may lose  subject identity. For example, it is easy for the network to learn a mapping to an image that corresponds to the target age and health state, but belongs to a different subject. An illustration is presented in Fig.~\ref{fig: illustration of identity loss}, where ageing trajectories of two subjects are shown. The task is to predict the brain image of subject 1 at age $a_{2}$ starting at age $a_1$, by learning a mapping from point A to point B. But there are no ground-truth data to ensure that we stay on the trajectory of subject 1. Instead, the training data contain brain images of age $a_2$ belonging to subject 2 (and other subjects). Using only $L_{GAN}$, the Generator may learn a mapping from A to C to fool the Discriminator, which will lose the identity of subject 1. To alleviate this and encourage the network to learn mappings along the trajectory (i.e.\ from A to B), we adopt:
\begin{equation}
\label{eq: L_ID}
\begin{split}
    L_{ID}=\E_{\mathbf{x}_{i}\sim \mathcal{X}_{i}, \hat{\mathbf{x}}_{o} \sim {\mathcal{X}}_{o}}{\lVert}\mathbf{x}_{i}-\hat{\mathbf{x}}_{o}{\rVert}_{1} \cdot e^{-\frac{|a_{o}-{a}_{i}|}{|a_{max}-a_{min}|}},
\end{split}
\end{equation}
where $\mathbf{x}_{i}$ is the input image of age $a_{i}$ and $\hat{\mathbf{x}}_{o}$ is the output image of age $a_{o}$ ($a_{o}>a_{i}$). The term $e^{-\frac{|a_{o}-a_{i}|}{|a_{max}-a_{min}|}}$ encourages ${\lVert}\mathbf{x}_{i}-\mathbf{\hat{x}}_{o}{\rVert}_{1}$ to positively correlate with the difference $|a_{o}-a_{i}|$. The health state is not involved in $L_{ID}$ as we do not aim to precisely model the ageing trajectory. Instead, $L_{ID}$ is used to encourage identity preservation by penalising major changes between images close in age, and to stabilise training. A more accurate ageing prediction, which is also correlated with health state, is achieved by the adversarial loss. An ablation study illustrating the critical role of $L_{ID}$ is included in Section~\ref{sec: ablation studies}.


\subsubsection{Self-reconstruction loss} 
We use a self-reconstruction loss,
\begin{equation}
\label{eq: L_self_rec}
\begin{split}
    L_{rec}=\E_{\mathbf{x}_{i}\sim \mathcal{X}_{i}, \hat{\mathbf{x}}_{o} \sim {\mathcal{X}}_{i}}{\lVert}\mathbf{x}_{i}-\hat{\mathbf{x}}_{o}{\rVert}_{1},
\end{split}
\end{equation}
to explicitly encourage that the output $\hat{\mathbf{x}}_{o}$ is a faithful reconstruction of the input $\mathbf{x}_{i}$ for the same age and health state.  Although $L_{rec}$ is similar to $L_{ID}$, their roles are different: $L_{ID}$ helps to preserve subject identity when generating aged images, while $L_{rec}$ encourages smooth progression via self-reconstruction.  An ablation study on $L_{rec}$ in Section~\ref{sec: ablation studies} shows the importance of stronger regularisation.\footnote{In our previous work~\citep{xia2019consistent}, Eq.~\ref{eq: L_ID} did not have the $a_{o} > a_{i}$ constraint and would randomly include the case of $a_{o} = a_{i}$ to encourage self-reconstruction. However, as shown in Section~\ref{sec: ablation studies}, we found that stronger regularisation is necessary.}

\section{Experimental setup}
\label{sec: experimenral setup}

\textbf{Datasets:} We use two datasets, as detailed below.

\textit{Cambridge Centre for Ageing and Neuroscience (Cam-CAN)}~\citep{camcan2015ageing} is a cross-sectional dataset containing normal subjects aged 18 to 88.  We split subjects into different age groups spanning 5 years. We randomly selected 38 volumes from each age group and used 30 for training and 8 for testing. To prevent data imbalance, we discarded subjects under 25 or over 85 years old, because there are underrepresented in the dataset. We use Cam-CAN to demonstrate consistent brain age synthesis across the whole lifespan. 
\textit{Alzheimer's Disease Neuroimaging Initiative (ADNI)}~\citep{petersen2010alzheimer} is a longitudinal dataset of subjects being cognitively normal (CN), mildly cognitive impaired (MCI) and with AD. We use ADNI to demonstrate brain image synthesis, conditioned on different health states. Since ADNI has longitudinal data, we used these data to quantitatively evaluate the quality of synthetically aged images. We chose 786 subjects as training (279 CN, 260 MCI, 247 AD), and 136 subjects as testing data (49 CN, 46 MCI, 41 AD). The age difference between baseline and followup images in the testing set is $2.93\pm 1.35$ years.

\textbf{Pre-processing:} All volumetric data are skull-stripped using DeepBrain\footnote{https://github.com/iitzco/deepbrain}, and linearly registered to MNI 152 space  using FSL-FLIRT~\citep{woolrich2009bayesian}. We normalise brain volumes by clipping the intensities to $[0,V_{99.5}]$, where $V_{99.5}$ is the 99.5\% largest intensity value within each volume, and then rescale the resulting intensities to the range $[-1,+1]$. Such intensity pre-processing also helps alleviate potential intensity harmonisation issues between datasets in a manner that creates no leakage (see footnote on section 5.2.3 why this is important). We select the middle 60 axial slices from each volume, and crop each slice to the size of $[208,160]$. During training, we only use \textit{cross-sectional} data, i.e.\ one subject only has one volume of a certain age. During testing, we use the longitudinal ADNI data covering more than 2 years, and discard data where images are severely misaligned due to registration errors.

\textbf{Benchmarks: }We compare with the following benchmarks\footnote{We also used the official implementation of~\citet{milana2017deep}; however, our experiments confirmed the poor image quality reported by the author.}:

\noindent 

\noindent \textit{Conditional GAN: } 
    We use a conditional image-to-image translation approach~\citep{mirza2014conditional} and train different Conditional GANs for transforming young images to different older age groups. Therefore, a single model of ours is compared with age-group specific Conditional GANs.
\\  
\noindent \textit{CycleGAN: } We use CycleGAN~\citep{zhu2017unpaired}, with two translation paths: from `young' to 'old' to `young', and from `old' to `young' to `old'. 
Similarly to Conditional GAN, we train several CycleGANs for different target age groups.

\noindent \textit{CAAE}: We compare with \citet{zhang2017age}, a recent paper for face ageing synthesis. We use the official implementation\footnote{https://zzutk.github.io/Face-ageing-CAAE/}, modified to fit our input image shape. This method used a Conditional Adversarial Autoencoder (CAAE) to perform face ageing synthesis by concatenating a one-hot age vector with the bottleneck vector. They divided age into discrete age groups.

\textbf{Implementation details: } The optimization function is: 
\begin{equation}
\begin{split}
    L = \stackunder{min}{G}\, \stackunder{max}{D} \lambda_{1} L_{GAN_{}}+\lambda_{2} L_{ID}+\lambda_{3} L_{rec},
\end{split}
\end{equation}
where $\lambda_{1}=1$, $\lambda_{2}=100$ and $\lambda_{3}=10$ are hyper-parameters used to balance each loss. The $\lambda$ parameters are chosen experimentally. We chose $\lambda_{2}$ as 100 following \citet{baumgartner2018visual} and \citet{xia2019consistent}, and $\lambda_{3}$ as a smaller value to put emphasis on synthesis rather than self-reconstruction.

To train our model, we divide subjects into a young group and an old group, and randomly draw a image $\mathbf{x}_{i}$ the young group and an image $\mathbf{y}_{o}$ from the old group to synthesise the aged image $\hat{\mathbf{x}}_{o}$ (of $\mathbf{x}_{i}$) with target age $a_{o}$ and health state $h_{o}$ (of those corresponding to $\mathbf{y}_{o}$). Here $\hat{\mathbf{x}}_{o}$ is the synthetically aged version of $\mathbf{x}_{i}$, and the target age $a_{o}$ and health state $h_{o}$ are the same as those of the selected old sample $\mathbf{y}_{o}$. Afterwards,  $\mathbf{y}_{o}$ and $\hat{\mathbf{x}}_{o}$ are fed into the discriminator as real and fake samples, respectively. Note that for all samples $a_{o}>a_{i}$, and $h_{o}$ could be different than $h_{i}$. Since Alzheimer's Disease is an irreversible neurodegenerative disease, we select samples where the input health status is not worse than the output health status. We train all methods for 600 epochs. We update the generator and discriminator iteratively~\citep{arjovsky2017wasserstein,goodfellow2014generative}. Since the discriminator of a Wasserstein GAN needs to be close to optimal during training, we update the discriminator for 5 iterations per generator update. Initially, for the first 20 epochs, we update the discriminator for 50 iterations per generator update. We use Keras~\citep{chollet2015keras} and train with Adam~\citep{kingma2015adam} with a learning rate of 0.0001 and decay of 0.0001. Code is available at \url{https://github.com/xiat0616/BrainAgeing}. 

\textbf{Evaluation metrics: } To evaluate the quality of synthetically aged images, we first use the longitudinal  data from ADNI dataset. We select follow-up studies covering $>$2 years to allow observable neurodegenerative changes to happen. We used standard definitions of \textit{mean squared error} (MSE),  \textit{peak signal-to-noise ratio} (PSNR) and \textit{structural similarity} (SSIM) of window length of 11 \citep{wang2003multiscale} to evaluate the closeness of the predicted images to the ground-truth.

\textit{Predicted age difference (PAD) as a metric:} 
Longitudinal data in ADNI only cover a short time span, i.e.\ the age difference between baseline and followup images is only several years. To assess output even when we do not have corresponding follow-up ground truth, we use a proxy metric of apparent age to evaluate image output. To develop our proxy metric, we first train a learning based age predictor to assess apparent brain age. We pre-train a VGG-like~\citep{simonyan2014very} network to predict age from brain images, then use this age predictor, $f_{pred}$, to estimate the apparent age of output images. 
To train this age predictor $f_{pred}$ we combine Cam-CAN and healthy (CN) ADNI training data to ensure good age coverage. On a held out testing set it achieves a MAE of $5.1\pm3.1$ years. When the held out dataset is restricted to ADNI healthy subjects alone, MAE is $3.9\pm2.8$ years.

We use the difference between the predicted and desired target age to assess how close the generated images are to the (desired) target age. Formally, our proxy metric \textit{predicted age difference} (PAD) is: 
\begin{equation}
\begin{split}
PAD=\E_{\hat{\mathbf{x}}_{o} \sim \mathcal{X}_{o}} | f_{pred} (\hat{\mathbf{x}}_{o})-a_{o}|,
\end{split}
\end{equation}
where $f_{pred}$ is the trained age predictor, $\hat{\mathbf{x}}_{o}$ is the synthetically aged image, and $a_{o}$ is the target age. Here we choose to measure the mean absolute error as we want to avoid  the neutralization of positive and negative errors. By adopting PAD, we have a quantitative metric to measure the quality of synthetic results in terms of age accuracy. Observe that PAD does not compare baseline and follow-up scans.  Given that the age predictor is only trained on healthy data it will estimate age on how normal brains will look like. Thus, it should capture when brain ageing acceleration occurs in AD, as others have demonstrated before us~\citep{cole2019brain}. This will increase PAD error when we synthesise with AD or MCI target health state, but given that we use PAD to compare between different methods this error should affect all methods.   With advances in brain ageing estimation~\citet{peng2021accurate} the fidelity of PAD will also increase. Here since we use PAD to compare across methods even a biased estimator is still a useful method of comparison.

\textbf{Statistics: } All results are obtained on testing sets, and we show average and standard deviation (std, as subscript on all tables), estimated by sample mean and variance on the testing set. We use \textbf{bold} font to denote the best performing method (for each metric) and an asterisk (*) to denote statistical significance. We use a paired t-test (at 5\% level assessed via permutations)  to test the null hypothesis that there is no difference between our methods and the best performing benchmark. 
\begin{figure*}[t]
    \centering
    \includegraphics[scale=0.56]{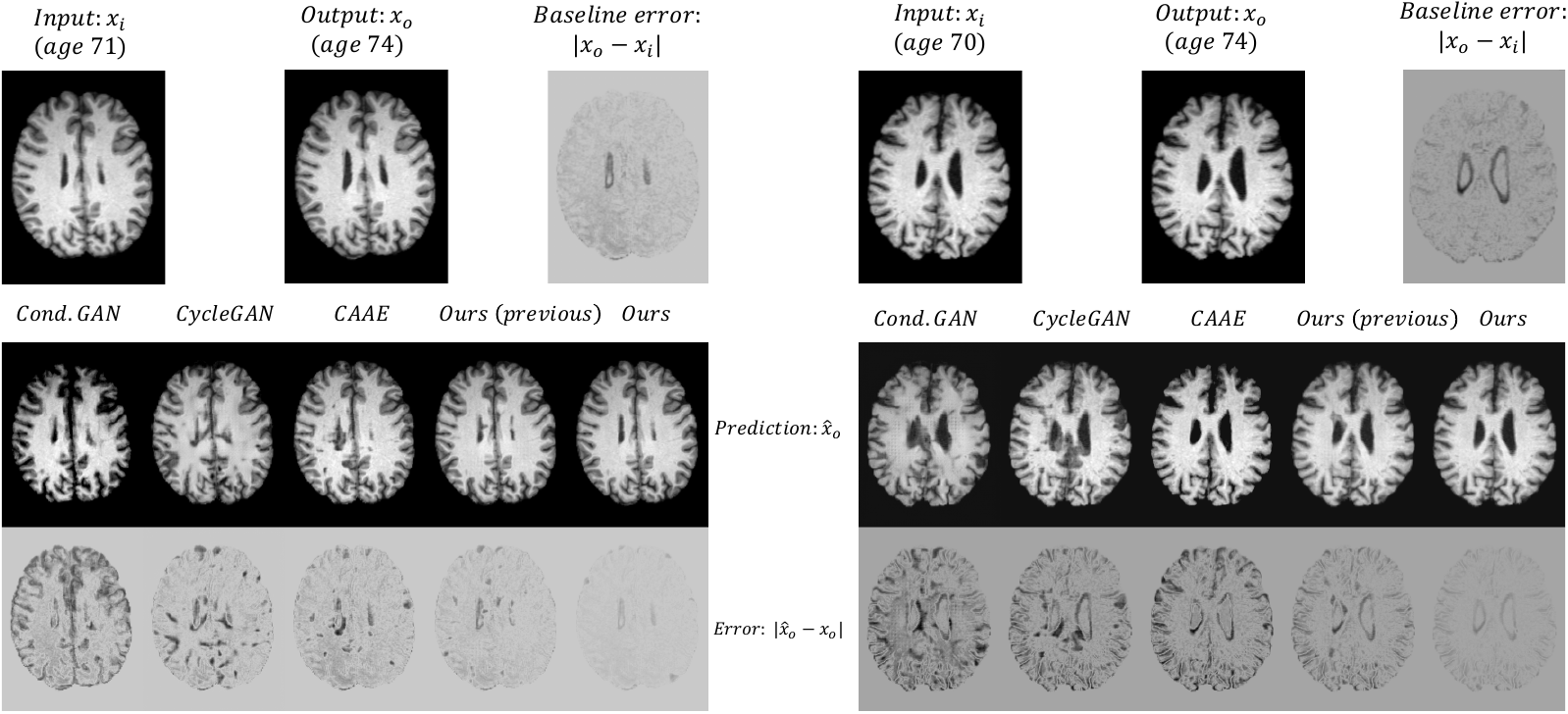}
    \caption{Example results of subjects with ground-truth follow-up studies. We predict output $\hat{x}_{o}$ from input $x_{i}$ using benchmarks and our method. We also show errors between the outputs and the ground-truths as $|\hat{x}_{o}-x_{o}|$. We can observe that our method achieves the most accurate results outperforming our previous method~\citep{xia2019consistent} and benchmarks. As a comparison, we also visualized the difference between inputs and ground-truth outputs as $|x_{o}-x_{i}|$. For more details see text.}  
    \label{fig:detailed longitudinal results}
\end{figure*}

\section{Results and discussion}
\label{sec: results and discussion}
We start by showing quantitative and qualitative results on ADNI  with detailed evaluation demonstrating quality of the generated images. We then train our model on Cam-CAN to show long-term brain ageing synthesis. We conclude with  ablation studies to illustrate the effect of design choices.

\begin{table*}[]
\centering
\caption{Quantitative evaluation on ADNI dataset (testing set) for several metrics. We report average and std (as subscript) with BOLD, * indicating best performance and statistical significance, respectively (see Section \ref{sec: experimenral setup}).}
\label{tab: longitudinal ADNI}
\begin{tabular}{l|lllllll}
\hline
                & SSIM & PSNR &  MSE & PAD & PAD (CN) & PAD (MCI) & PAD (AD) \\ \hline
Naïve baseline & $0.71_{\pm0.09}$ & $22.1_{\pm3.3}$ & $0.097_{\pm0.013}$& $7.2_{\pm 3.9}$ & $6.3_{\pm 3.8}$ & $6.8_{\pm 3.9}$ & $8.7_{\pm 4.0}$\\
Cond. GAN &   $0.39_{\pm 0.08}$     & $14.2_{\pm 3.5}$     &  $0.202_{\pm 0.012}$ &  $9.5_{\pm 4.7}$ & $8.7_{\pm 4.8}$     &  $9.1_{\pm 4.7}$ &  $10.9_{\pm 4.7}$     \\
CycleGAN        &    $0.46_{\pm 0.07}$    &   $16.3_{\pm 3.3}$      &    $0.193_{\pm 0.008}$ &    $9.7_{\pm 5.1}$  & $8.9_{\pm 4.9}$  &$9.4_{\pm 5.2}$  &$11.0_{\pm 5.2}$     \\
CAAE   &   $0.64_{\pm 0.07}$   & $20.3_{\pm 2.9}$      &     $0.114_{\pm 0.011}$   & $5.4_{\pm 4.5}$ & $4.4_{\pm 4.3}$ & $5.1_{\pm 4.4}$ & $6.9_{\pm 4.7}$ \\
Ours-previous  &  $0.73_{\pm 0.06}$    &   $23.3_{\pm 2.2}$   &   $0.081_{\pm 0.009}$  & $5.0_{\pm 3.7}$ & $4.0_{\pm 3.5}$ & $4.6_{\pm 3.6}$ & $6.6_{\pm 4.0}$   \\ \hline
Ours            &  $\mathbf{0.79^{*}_{\pm 0.06}}$    &  $ \mathbf{26.1^{*}_{\pm 2.6}}$    &   $\mathbf{0.042^{*}_{\pm 0.006}}$   & $\mathbf{4.2^{*}_{\pm 3.9}}$  & $\mathbf{3.1^{*}_{\pm 3.6}}$ &
$\mathbf{3.9^{*}_{\pm 3.8}}$ &
$\mathbf{5.9^{*}_{\pm 4.2}}$ \\ \hline
\end{tabular}
\end{table*}

\subsection{Brain ageing synthesis on different health states (ADNI)}
\label{sec: synthesis under different health state}
In this section, we train and evaluate our model on ADNI dataset, which contains CN, MCI and AD subjects. Our model is trained only on \textit{cross-sectional} data. The results and discussions are detailed below.

\subsubsection{Quantitative results}
The quantitative results are shown in Table~\ref{tab: longitudinal ADNI}, employing the metrics defined in Section~\ref{sec: experimenral setup}.  For ADNI we also obtained a non-learned naïve baseline that simply calculates performance comparing ground-truth baseline and follow-up images. The naïve baseline result is obtained by subtracting from the followup the baseline (input) image. We involve this non-learned baseline as a lower bound to check if the proposed algorithm synthesises 
images that are closer to the follow-up than the baseline images or not. As reported in Section 4, the average age prediction error (MAE) of the age predictor on the ADNI testing data is 3.9 years. Estimating PAD separately for CN, MCI and AD testing subjects (see Table~\ref{tab: longitudinal ADNI}) shows that the best PAD results are obtained on healthy (CN) data. This is expected as the age predictor used to estimate PAD it is trained on healthy data only.   However, this bias affects all methods, and thus still allows comparisons between them. Indeed, we can observe that our method achieves the best results in all metrics, with second best being the previous (more simple incarnation)~\citep{xia2019consistent} of the proposed model.  Embedding health state improves performance, because it permits the method to learn an ageing function specific for each state as opposed to the one learned by the method in~\citet{xia2019consistent}.  The other benchmarks achieve a lower performance compared to the baseline. The next best results are achieved by CAAE~\citep{zhang2017age}, where age is divided into 10 age groups and represented by a one-hot vector. To generate the aged images at the target age (the age of the follow-up studies), we use the age group to which the target age belongs, i.e.\ if the target age is 76, then we choose the age group of age 75-78. We see the benefits of encoding age into ordinal vectors, where the difference between two vectors positively correlates with the difference between two ages in a finely-grained fashion. CycleGAN and Conditional GAN achieve the poorest results unsurprisingly, since conditioning here happens explicitly by training separate models according to different age groups. 

\subsubsection{Qualitative results}
\begin{figure*}[t!]
    \centering
    \includegraphics[scale=0.56]{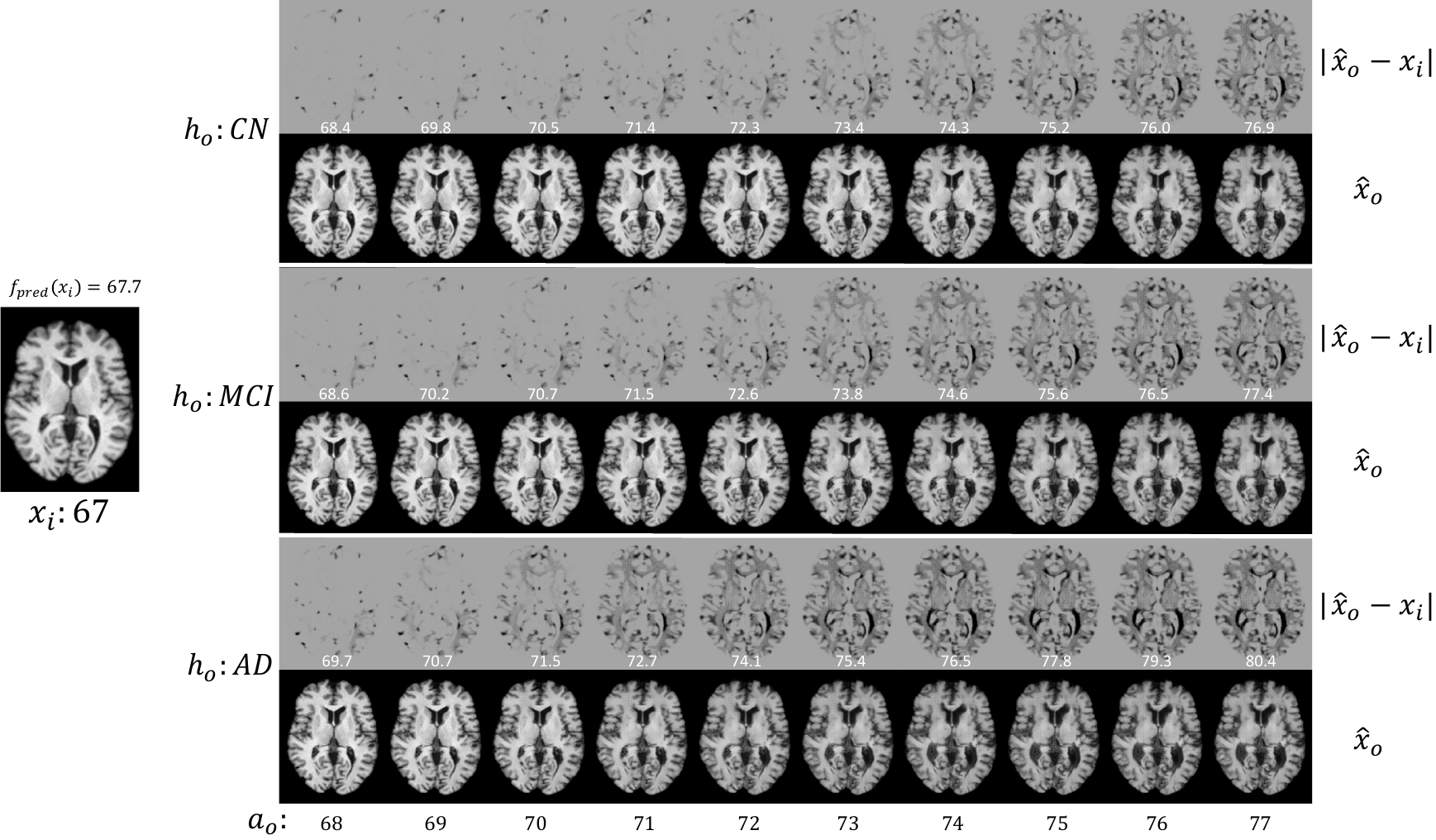}
    \caption{Brain ageing progression for a healthy (CN) subject $\mathbf{x}_{i}$ (at age 67) from ADNI dataset. We synthesise the aged images $\hat{\mathbf{x}}_{o}$ at different target ages $a_{o}$ on different health states $h_{o}$: CN, MCI and AD, respectively. We also visualise the difference between $\mathbf{x}_{i}$ and $\hat{\mathbf{x}}_{o}$,  $|\hat{\mathbf{x}}_{o}-\mathbf{x}_{i}|$, and show the predicted (apparent) ages of $\hat{\mathbf{x}}_{o}$ as obtained by our pre-trained age predictor (white text overlaid on each difference image). 
    For more details see text. } 
    \label{fig:ADNI different AD diagnosis}
\end{figure*}
Visual examples on two images from ADNI, are shown in Fig.~\ref{fig:detailed longitudinal results}. For both examples, our method generates most accurate predictions, followed by our previous method ~\cite{xia2019consistent}, offering visual evidence to the observations above. The third best results are achieved by CAAE, where we can see more errors between prediction $\hat{x_{o}}$ and ground-truth $x_{o}$. CycleGAN and Conditional GAN produced the poorest output images, with observable structural differences from ground-truth, indicating loss of subject identity. We can also observe that the brain ventricle is enlarged in our results and  the difference between $x_i$ and $x_o$ is reduced, which is consistent with known knowledge that ventricle increases during ageing.

Furthermore, we show visual results of the same subject at different target health states $h_{o}$, in Fig.~\ref{fig:ADNI different AD diagnosis}. We observe that for all $h_{o}$, the brain changes gradually as age ($a_{o}$) increases. However, the ageing rate varies based on health state ($h_{o}$). Specifically, when $h_{o}$ is CN, ageing is slower than that of MCI and AD, as one would expect; when $h_{o}$ is AD, ageing changes accelerate. We also report the estimated ages of these synthetic images as predicted by $f_{pred}$. While these results show one instance, we synthesised aged images of different health status from 49 ADNI test set CN subjects, with target ages 10 years older than the original age. We then used $f_{pred}$ to estimate the age of these synthetic images. We find that on average, synthetic AD images are $4.9\pm 2.3$ years older than the target age, whereas synthetic MCI and CN images are $1.8\pm 2.0$ and $1.5\pm 2.1$ years older than the target age, respectively.  These observations are consistent with prior findings that AD accelerates  brain ageing~\citep{petersen2010alzheimer}. We also observe that the gray/white matter contrast decreases as age increases, which is consistent with existing findings~\citep{westlye2009increased,farokhian2017age}. 

\subsection{Does our model capture realistic morphological changes of ageing and disease?} 
Here we want to assess whether our model captures known ageing-related brain degeneration. It is known that brain ageing is related to gray matter reduction in middle temporal gyrus (MTG)~\citep{guo2014grey,sullivan1995age}. We wanted to assess whether synthetic volumes could act as drop-in replacements of ground-truth follow-up in assessing MTG gray matter volume change. We focus here on the MTG as this is well covered by the range of slices we use to train our synthesis method.  Before we proceed we first illustrate that we can synthesise 3D volumes slice-by-slice, and then show that our model can capture realistic morphological changes.

\subsubsection{Volume synthesis by stacking 2D slices}
\label{sec: volume synthesis}
\begin{figure}[t]
    \centering
    \includegraphics[scale=0.4]{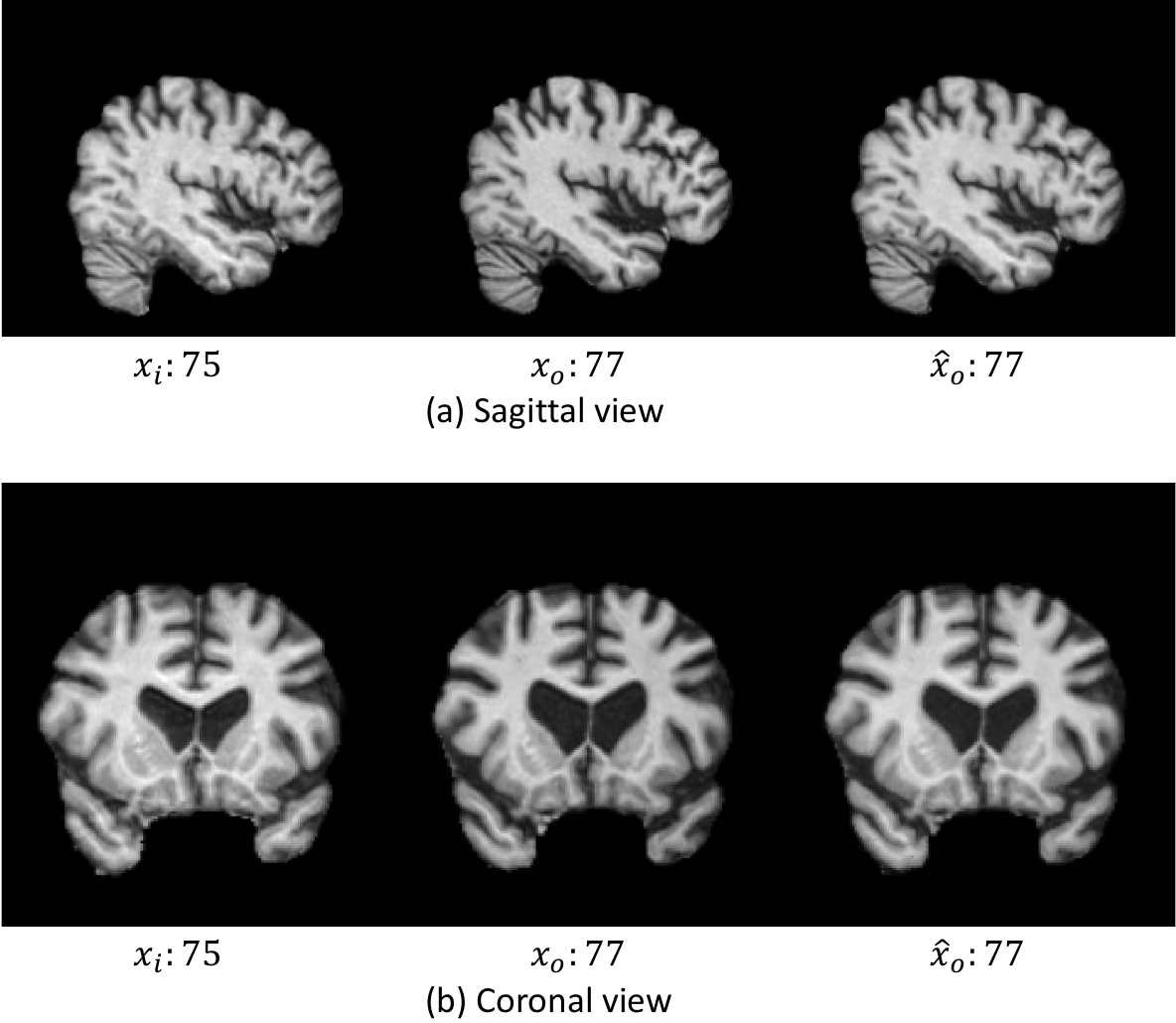}
    \caption{Example results of a synthetic 3D volume $\hat{\mathbf{x}}_{o}$ in sagittal view (top) and coronal view (bottom) from ADNI dataset. Here we construct the 3D volume by stacking the 2D synthetic axial slices of our model. From left to right are slices from a baseline volume $\mathbf{x}_{i}$, the corresponding follow-up volume $\mathbf{x}_{o}$, and the stacked synthetic volume $\hat{\mathbf{x}}_{o}$. }
    \label{fig: 3D views}
\end{figure}

We show that, even with our 2D model, we can still produce 3D volumes that show consistency. We applied our model on 2D axial slices and obtained a 3D volume by stacking the synthetic slices. An example result of a stacked synthetic 3D volume in sagittal and coronal views is shown in Figure~\ref{fig: 3D views}.  Compared to the respective ground-truth from the same subject, we observe that both sagittal and coronal views of the synthetic volume look realistic and are close to the follow-up images. Note here that our model is trained only on 2D axial slices, for which we chose middle 60 slices from each volume. Our model uses a residual connection and thus makes minimal changes to the regions affected by age instead of synthesising the whole brain image. This helps preserve details and continuity across slices. These results illustrate that we can produce 3D volumes that maintain consistency in different views.


\subsubsection{Do we capture morphological changes?}
\begin{figure}[b]
    \centering
    \includegraphics[scale=0.28]{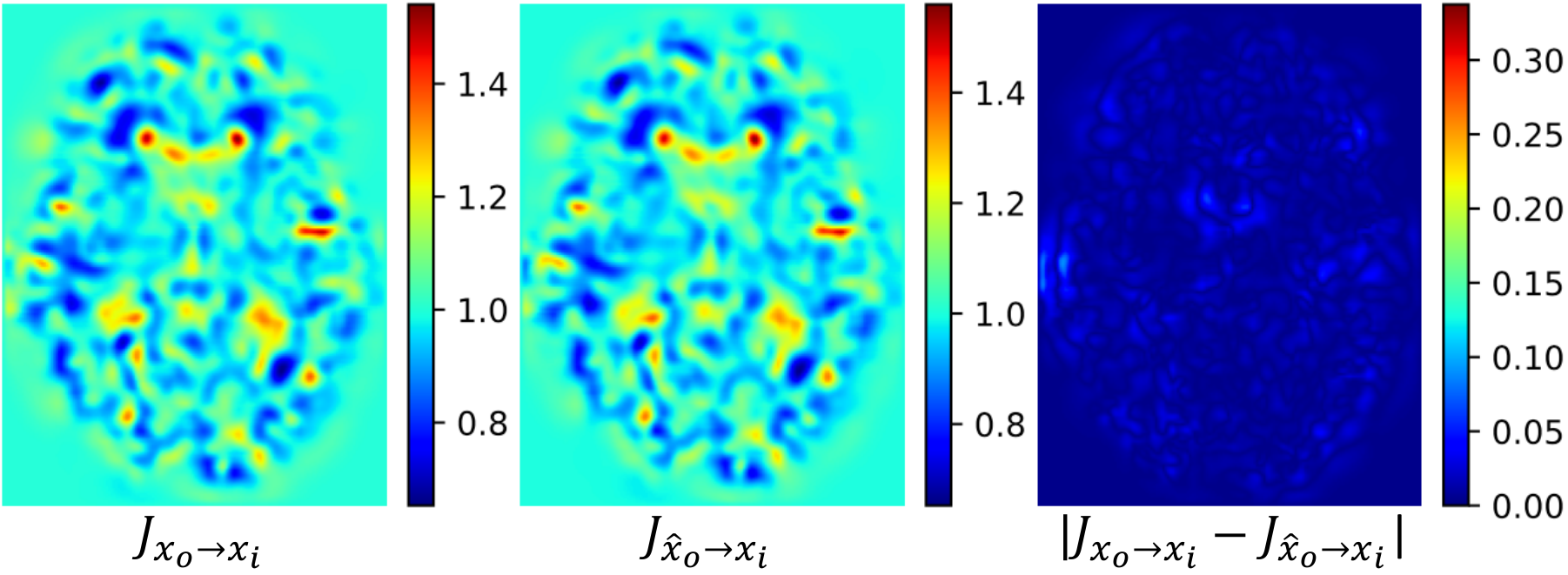}
    \caption{ An example of Jacobian determinant maps for a subject. From left to right are the Jacobian determinant maps $\mathbf{J}_{\mathbf{x}_{o}\rightarrow \mathbf{x}_{i}}$, $\mathbf{J}_{\mathbf{\hat{x}}_{o} \rightarrow \mathbf{x}_{i}}$, and the error map between them: $|\mathbf{J}_{\mathbf{x}_{o}\rightarrow \mathbf{x}_{i}}-\mathbf{J}_{\mathbf{\hat{x}}_{o} \rightarrow \mathbf{x}_{i}}|$.
}
    \label{fig: jacobian}
\end{figure}

We use an $\ell_{1}$ loss to restrict (in pixel space) the amount of change between input and output images.  This is computationally efficient, but to show that it also restricts deformations, we measure the deformation between input (baseline) and synthetic or ground-truth follow-up images in ADNI. We obtain for each subject the baseline image $\mathbf{x}_{i}$, the follow-up image $\mathbf{x}_{o}$ and the synthetic image $\hat{\mathbf{x}}_{o}$, respectively. We first rigidly register $\mathbf{x}_{o}$ to $\mathbf{x}_{i}$ using Advanced Normalization Tools (ANTs)~\citep{avants2008symmetric} rigid transformation. Then we non-rigidly register $\mathbf{x}_{o}$ to  $\mathbf{x}_{i}$ and obtain the Jacobian determinant map $\mathbf{J}_{\mathbf{x}_{o}\rightarrow \mathbf{x}_{i}}$ that describes the transformation from $\mathbf{x}_{o}$  to $\mathbf{x}_{i}$, using ANTs "SyN" transformation~\citep{avants2008symmetric}. Similarly, we obtain $\mathbf{J}_{\mathbf{\hat{x}}_{o} \rightarrow \mathbf{x}_{i}}$ that describes the non-linear transformation from $\mathbf{\hat{x}}_{o}$ to $\mathbf{x}_{i}$. Fig.~\ref{fig: jacobian} shows an example of the Jacobian  maps for one subject.

From Fig.~\ref{fig: jacobian}, we observe that $\mathbf{J}_{\mathbf{\hat{x}}_{o} \rightarrow \mathbf{x}_{i}}$ is close to $\mathbf{J}_{\mathbf{x}_{o}\rightarrow \mathbf{x}_{i}}$. To quantify their difference, we calculate the mean relative error between the Jacobian determinant maps, defined as:
\begin{equation}
    E=\E_{\mathbf{x}_{i}\sim \mathcal{X}_{i}, \mathbf{x}_{o} \sim \mathcal{X}_{o}, \hat{\mathbf{x}}_{o} \sim {\hat{\mathcal{X}}_{o}}} \frac{\lVert \mathbf{J}_{\mathbf{x}_{o}\rightarrow \mathbf{x}_{i}}-\mathbf{J}_{\mathbf{\hat{x}}_{o} \rightarrow \mathbf{x}_{i}} \rVert_{1}}{\lVert \mathbf{J}_{\mathbf{x}_{o}\rightarrow \mathbf{x}_{i}} \rVert_{1}},
\end{equation}
\noindent where $\lVert . \rVert_{1}$ is 1-norm of matrices. We find the mean relative error to be $3.49\%$ on the testing set of 136 images. Similarly, we perform the same evaluations for the results of Conditional GAN, CycleGAN, CAAE and our previous method, and find the mean relative errors to be $9.87\%$, $8.76\%$, $5.91\%$ and $4.43\%$, respectively.
Both qualitative and quantitative results suggest that synthetically aged images capture realistic morphological changes of the brain ageing process.

\subsubsection{Measuring middle temporal gyrus (MTG) gray matter atrophy.}
\label{sec:mtg}

We further evaluate the quality of the synthetic results by assessing if they can act as a drop-in replacement to real data in a simple study of brain atrophy. We performed ageing synthesis with our model on 136 ADNI testing subjects, such that for each subject we have: a baseline image $\mathbf{x}_i$; a real follow-up image $\mathbf{x}_{o}$; and a synthetic image $\hat{\mathbf{x}}_{o}$ of the same target age and health state as of $\mathbf{x}_{o}$. We then assembled volumes by stacking 2D images. Then we affinely registered both $\mathbf{x}_{o}$ and $\hat{\mathbf{x}}_{o}$ and the Human-Brainnetome based on Connectivity Profiles (HCP) atlas~\citep{fan2016human} to $\mathbf{x}_i$. After that, we obtained the MTG segmentation of $\mathbf{x}_i$, $\mathbf{x}_{o}$ and $\hat{\mathbf{x}}_{o}$ by means of label propagation from HCP using the deformation fields. Then we obtained the gray matter segmentation of $\mathbf{x}_i$ using FSL-FAST~\citep{zhang2001segmentation}. The gray matter segmentation of $x_{o}$ and $\hat{\mathbf{x}}_{o}$ was subsequently obtained by non-linearly registering $\mathbf{x}_{o}$ and $\hat{\mathbf{x}}_{o}$ to  $\mathbf{x}_i$ and propagating anatomical labels using ANTs~\citep{avants2008symmetric}. These steps yield the MTG gray matter volume of $\mathbf{x}_{i}$, $\mathbf{x}_{o}$ and $\hat{\mathbf{x}}_{o}$, termed as $\mathbf{V}_{base}$, $\mathbf{V}_{fol}$, and $\mathbf{V}_{syn}$, respectively. Then, we calculate the relative change between $\mathbf{V}_{base}$ and $\mathbf{V}_{fol}$ as $RC_{real}=\frac{\mathbf{V}_{fol}-\mathbf{V}_{base}}{\mathbf{V}_{base}}$, and the relative change for synthetic data as $RC_{syn}=\frac{\mathbf{V}_{syn}-\mathbf{V}_{base}}{\mathbf{V}_{base}}$. We repeat this for several subjects in three patient type groups, i.e.\ CN (49), MCI (46) and AD (41). 

We expect, following \citet{guo2014grey} and \citet{sullivan1995age}, to see a statistical relationship between patient type and $RC_{real}$ when assessed with a one-way analysis of variance (ANOVA). If a similar relationship is shown also with synthetic data $RC_{syn}$, 
it will demonstrate that for this statistical test,  our synthetic data can act as a drop-in replacement to real data, and as such have high quality and fidelity.   

The results are summarised in Table~\ref{tab: MTG gray}, where we report also the F-statistic of the omnibus one-way ANOVA test. We observe that MTG gray matter volume reduces in both real and synthetic volumes. This indicates that our synthetic results achieve good quality and similar statistical conclusions can be drawn  using real  or synthetic data in this simple atrophy study.

\begin{table}[t]
\centering
\caption{Analysis of MTG gray matter relative change between baseline and follow-up real or synthetic.  Mean and std are reported as well as the corresponding F-statistic of a one-way ANOVA test (between relative change and patient type), with asterisk indicating significance ($p<0.05$).}
\begin{tabular}{r|ccc}
\hline
\multicolumn{1}{l|}{} & Relative change & F-statistic \\ \hline
real ($RC_{real}$)          & $-0.071_{\pm 0.0096}$      & $4.008^{*}$   \\ \hline
synthetic ($RC_{syn}$)            & $-0.083_{\pm 0.0099}$      & $4.539^{*}$   \\ \hline
\end{tabular}
\label{tab: MTG gray}
\end{table}

\subsection{Long term brain ageing synthesis}
\begin{figure*}[t]
    \centering
    \includegraphics[scale=0.57]{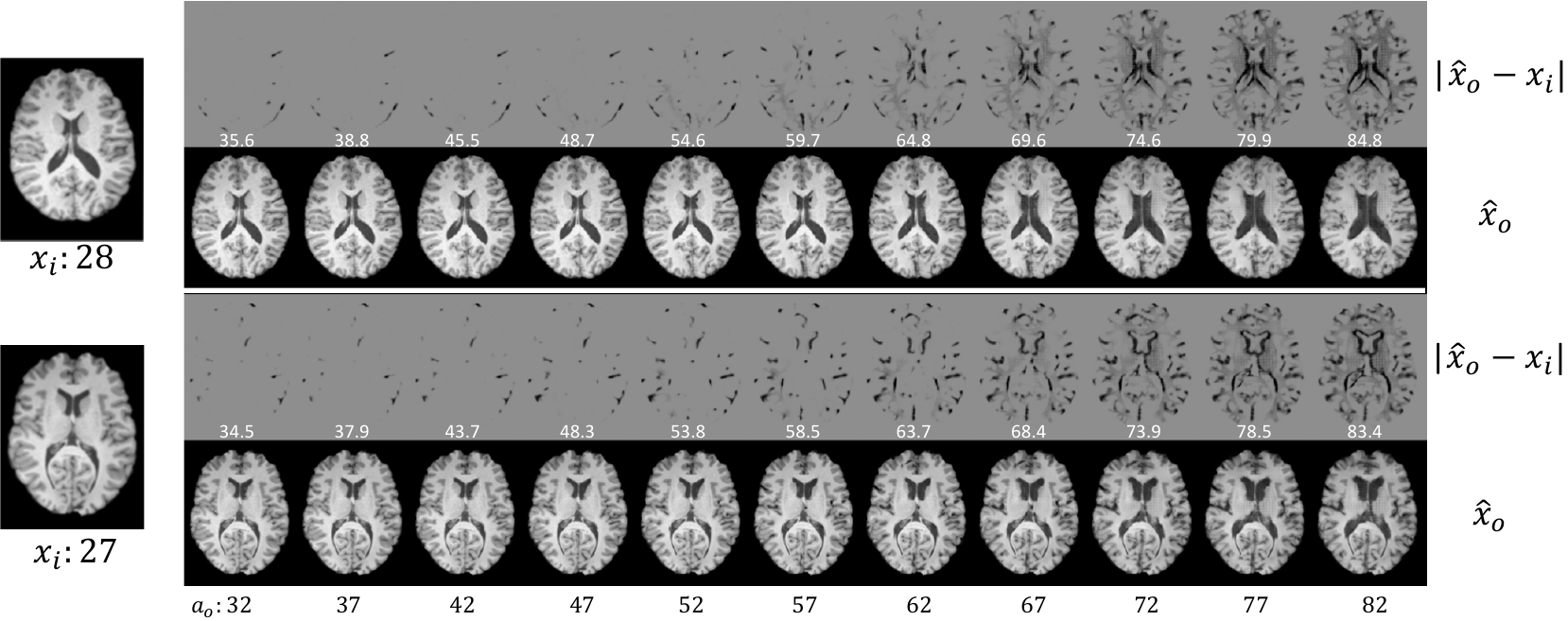}
    \caption{Long-term brain ageing synthesis on Cam-CAN dataset. We synthesise the aged images $\hat{\mathbf{x}}_{o}$ at different target ages $a_{o}$ and show the difference between input images $\mathbf{x}_{i}$ and $\hat{\mathbf{x}}_{o}$,  $|\hat{\mathbf{x}}_{o}- \mathbf{x}_{i}|$, and show the predicted (apparent) ages of $\hat{\mathbf{x}}_{o}$ as obtained by our pre-trained age predictor (white text overlaid on each difference image). Note here $\mathbf{x}_i$: N means an input image at age N. For more details see text.}
    \label{fig: results_cam_can_life_long}
\end{figure*}
\begin{figure*}[t]
    \centering
    \includegraphics[scale=0.53]{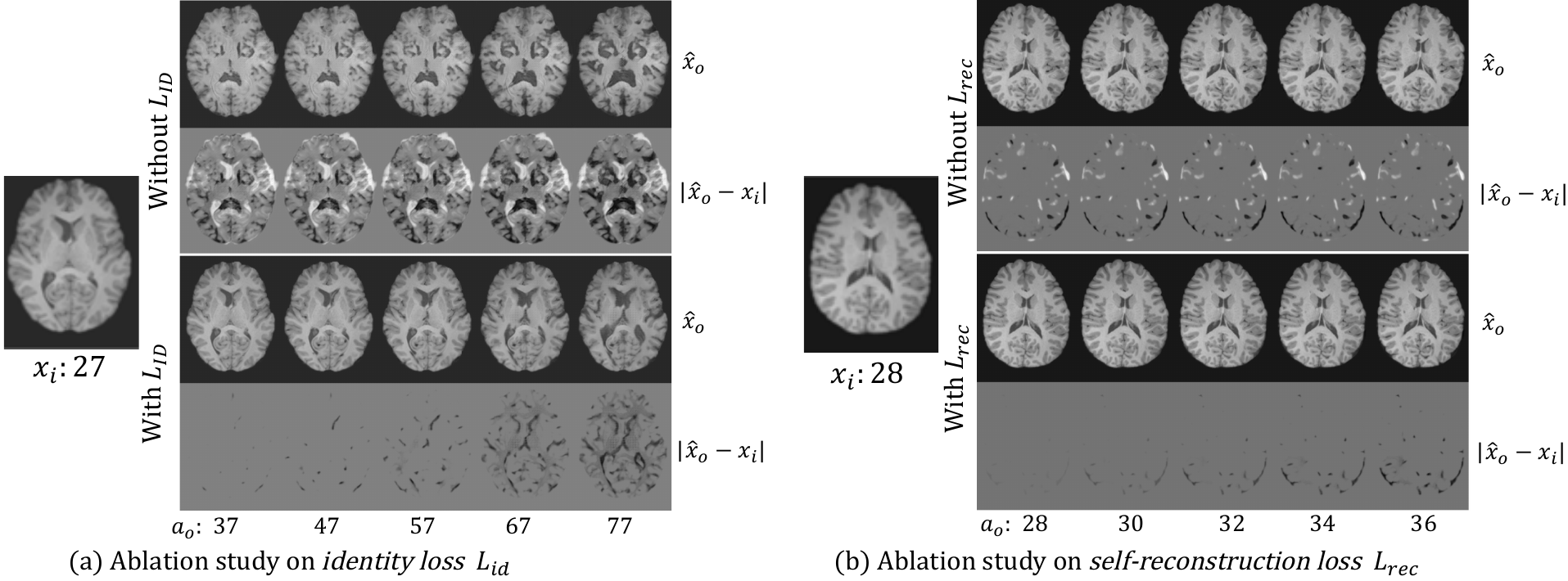}
    \caption{Ablation studies for loss components. \textbf{Left:} ablation study of $L_{ID}$. Top row shows that without $L_{ID}$, the network can lose the subject identity. Bottom row shows that the use of $L_{ID}$ can enforce the preservation of subject identity, such that the changes as ages are smooth and consistent. \textbf{Right:} ablation study on $L_{rec}$. When $L_{rec}$ is not used (top two rows), there are sudden changes at the beginning of ageing progression simulation (even at the original age), which hinders the preservation of subject identity. In contrast, when $L_{rec}$ is used (bottom two rows), the ageing progression is smoother, which demonstrates better identity preservation.  Note here $x_i$: N means an input image at age N.  }
    \label{fig: ablation studies for losses}
\end{figure*}
In this section, we want to see how our model performs in long term brain ageing synthesis. As ADNI dataset only covers old subjects, we use Cam-CAN dataset which contains subjects of all ages.
We train our model with Cam-CAN dataset where no longitudinal data are available, but evaluate it on the longitudinal part of ADNI to assess the generalisation performance of our model when trained on one dataset and tested on another.
%

%
%
%
\begin{figure*}[t]
    \centering
    \includegraphics[scale=0.55]{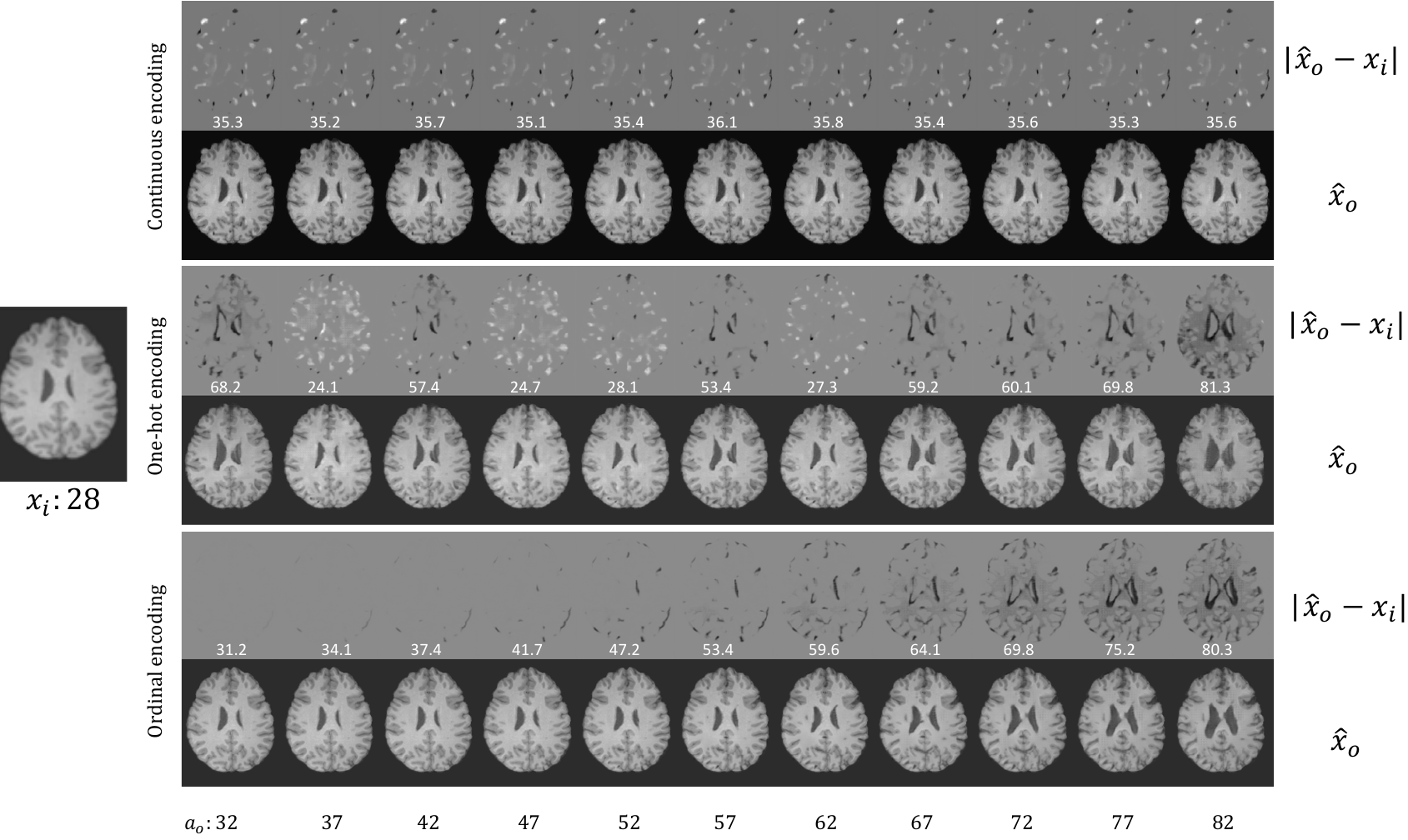}
    \caption{Example results for \textit{continuous}, \textit{one-hot} and \textit{ordinal} encoding on the Cam-CAN dataset for an image ($\mathbf{x}_i$) of a 28 year old subject. We synthesise aged images $\hat{\mathbf{x}}_{o}$ at different target ages $a_{o}$. We also show the difference between   $\mathbf{x}_{i}$ and $\hat{\mathbf{x}_{o}}$, $|\hat{\mathbf{x}}_{o} -\mathbf{x_{i}}|$, and report estimated age (white text overlaid at the bottom of each difference image). The proposed ordinal encoding shows consistent and progressive changes. }
    \label{fig: ablation study for one-hot}
\end{figure*}

\subsubsection{Qualitative results}
In Fig.~\ref{fig: results_cam_can_life_long}, we demonstrate the simulated brain ageing process throughout the whole lifespan, where the input images are two young subjects from Cam-CAN.
We observe that the output gradually changes as $a_{o}$ increases, with ventricular enlargement and brain tissue reduction. This pattern is consistent with previous studies~\citep{good2001voxel,mietchen2009computational}, showing that our method learns to  synthesise the ageing brain throughout the lifespan even trained on cross-sectional data.\footnote{We observe checkerboard artefacts near the ventricles after target age 67. Such artefacts are a known problem in computer vision and mostly likely due to the use of deconvolutional layers in the decoder ~\citep{oramas2018visual}.} 
 Fig.~\ref{fig: results_cam_can_life_long} offers only only a qualitative visualization to show the potential of life-time simulation. We cannot quantitatively evaluate the quality of these synthetic images due to the lack of longitudinal data in Cam-CAN. However, both the previous section on ADNI where we train and test on ADNI, and the next section, where we use longitudinal ADNI as testing set we but train on Cam-CAN data, offer considerable quantitative experiments. 

\subsubsection{Quantitative results (generalisation performance on ADNI)}
To evaluate how accurate our longitudinal estimation is, even when training with cross sectional data from \emph{another} dataset, we train a model on Cam-CAN and evaluate it on ADNI.  
We use the longitudinal portion of ADNI data, and specifically only the CN cohort, to demonstrate  generalisation performance.\footnote{We purposely do not use any intensity harmonisation that uses both datasets, e.g.\ histogram matching. Such methods will leak information from ADNI to Cam-CAN. Any leakage would skew (to our favour) the generalisation ability which we want to avoid. Thus, our experiments also indirectly evaluate how design choices (e.g.\ using a residual connection in the generator) help with differences in intensities between datasets. } Given an image of ADNI we use our Cam-CAN trained model to predict an output at the same age as the real follow up image.  We compare our prediction with the ground truth follow up image (in the ADNI dataset).
The results are shown in Table~\ref{tab: longitudinal Cam-CAN}. We observe that though our model is trained and evaluated on different datasets, it still achieves comparable results with those of Table~\ref{tab: longitudinal ADNI} and outperforms benchmarks. 

\begin{table}[t]
\centering
\caption{Quantitative evaluation of methods trained on Cam-CAN and evaluated on ADNI.}
\label{tab: longitudinal Cam-CAN}
\begin{tabular}{l|lllll}
\hline
                & SSIM & PSNR & MSE  & PAD \\ \hline
Cond. GAN &  $0.38_{\pm 0.12}$    &    $13.9_{\pm 4.2}$      & $0.221_{\pm 0.021}$     &   $11.3_{\pm 5.6}$  \\
CycleGAN        &  $0.42_{\pm 0.09}$    & $14.4_{\pm 3.8}$          &  $0.212_{\pm 0.016}$   &  $10.2_{\pm 5.5}$    \\
CAAE &  $0.59_{\pm0.10}$    &  $19.3_{\pm 3.9}$         &  $0.121_{\pm 0.012}$   & $5.9_{\pm 4.7}$    \\
Ours-previous &   $0.68_{\pm 0.08}$   & $22.7_{\pm 2.8}$           &  $0.095_{\pm 0.014}$    & $5.3_{\pm 3.8}$     \\ \hline
Ours     &   $\mathbf{0.74^{*}_{\pm 0.08}}$    &  $\mathbf{24.2^{*}_{\pm 2.7}}$      &    $\mathbf{0.043^{*}_{\pm 0.009}}$       &     $\mathbf{5.0_{\pm 3.6}}$     \\ \hline
\end{tabular}
\end{table}

\subsection{Ablation studies}
\label{sec: ablation studies}
We ablate loss components, explore different conditioning mechanisms, and explore latent space dimensions.

\subsubsection{Effect of loss components}
We demonstrate the effect of $L_{ID}$ and $L_{rec}$ by assessing the model performance when each component is removed. In Table~\ref{tab: ablation study of losses} we show quantitative results on ADNI dataset. In Fig.~\ref{fig: ablation studies for losses} we illustrate qualitative results on Cam-CAN dataset to visualise the effect. We can observe that the best results are achieved when all loss components are used. Specifically, without $L_{ID}$, the synthetic images lost subject identity severely throughout the whole progression, i.e.\ the output image appears to come from a different subject; without $L_{rec}$, output images suffer from sudden changes at the beginning of progression, even when $a_{o}=a_{i}$. Both quantitative and qualitative results show that the design of $L_{ID}$ and $L_{rec}$ improves preservation of subject identity and enables more accurate brain ageing simulation.
\begin{table}[b]
\caption{Ablations on using different combinations of cost functions.}
\label{tab: ablation study of losses}

\begin{tabular}{l|lllll}
\hline
                                       & SSIM & PSNR & MSE  \\ \hline
$L_{GAN_{}}$             &          $0.55_{\pm 0.14}$     &  $18.4_{\pm 3.7}$    &   $0.132_{\pm 0.013}$             \\
$L_{GAN_{}}+L_{rec}$                     &   $0.62_{\pm 0.12}$   &  $19.6_{\pm 3.2}$    &  $0.089_{\pm 0.014}$        \\
$L_{GAN_{}}+L_{ID}$                    &   $0.74_{\pm 0.07}$   & $24.3_{\pm 2.5}$      &  $0.074_{\pm 0.010}$        \\ \hline
$L_{GAN_{}}+L_{ID}+L_{rec}$             &  $\mathbf{0.79^{*}_{\pm 0.08}}$    & $\mathbf{26.1^{*}_{\pm 2.6}}$     &   $\mathbf{0.042^{*}_{\pm 0.006}}$       \\ \hline

\end{tabular}
\end{table}

\subsubsection{Effect of different embedding mechanisms}
We investigate the effect of different embedding mechanisms. Our embedding mechanism is described in Section~\ref{sec: proposed approach}. We considered to encode age as a normalized \textit{continuous} value (between 0 and 1) or using a \textit{one-hot} vector, which was then concatenated with the latent vector at the bottleneck.
The qualitative results are shown in Fig.~\ref{fig: ablation study for one-hot}. We can see that when age is represented as a normalized \textit{continuous} value, this is ignored by the network, thus generating similar images regardless of changes in target age $a_{o}$.   When we use \textit{one-hot} vectors to encode age, the network still generates realistic  images, but the ageing progression is not consistent, i.e.\ synthetic brains appear to have ventricle enlarging or shrinking in random fashion across age. In contrast, with \textit{ordinal encoding}, the model simulates the ageing process consistently. This observation is confirmed by the estimated ages of the output images by $f_{pred}$. 

We also compare with an embedding strategy where we concatenate $\mathbf{h}_{o}$, $\mathbf{a}_{d}$ and the bottleneck latent vector $\mathbf{v}_{1}$ together, and the concatenated vector is processed by the Decoder to generate the output image. We refer to this embedding strategy as $\mathit{concat_{all}}$. Results on ADNI are shown in Table~\ref{tab: ablation study of conditioning}. We found with $\mathit{concat_{all}}$, the network tends to ignore the health state vector $\mathbf{h}_{o}$ and only use $\mathbf{a}_{d}$. This can be caused by the dimensional imbalance between $\mathbf{h}_{o}$ ($2\times1$) and $\mathbf{a}_{d}$ ($100\times1)$.  When \textit{one-hot encoding} is used, performance deteriorates even more.

\begin{table}[t]
\centering
\caption{Quantitative results of  different embedding mechanisms.}
\label{tab: ablation study of conditioning}

\begin{tabular}{l|lllll}
\hline
                     & SSIM & PSNR & MSE  &PAD\\ \hline
One-hot              &   $0.54_{\pm 0.14}$   &  $17.3_{\pm 3.8}$    &  $0.177_{\pm 0.014}$ &$9.7_{\pm 4.9 }$        \\
concat\textsubscript{all}   &   $0.74_{\pm 0.09}$   &    $23.9_{\pm 2.9}$  & $0.065_{\pm 0.011}$   & $5.2_{\pm 3.9}$      \\
\hline
Ours & $\mathbf{0.79^{*}_{\pm 0.08}}$    & $\mathbf{26.1^{*}_{\pm 2.6}}$     &   $\mathbf{0.042^{*}_{\pm 0.006}}$  &  $\mathbf{5.0_{\pm 3.6}}$        \\
 \hline

\end{tabular}
\end{table}

\subsubsection{Effect of latent space dimension}
We explored whether latent dimension affects performance. We altered the length of the latent vector ($\mathbf{v}_2$) from $130 \times 1$ to twice smaller/larger and compared the corresponding models on ADNI. Our findings are shown in Table~\ref{tab: ablation study of latent space}. We find that our choice ($130 \times 1$) achieved the best results. It appears that too small is not enough to represent image information well, and too large can cause dimension imbalance. 

\subsubsection{Comparison with longitudinal model}
To compare our method with models that use longitudinal data\footnote{ \cite{ravi2019degenerative} also synthesise subject-specific aged brain images based on longitudinal data relyin on complex biological constraints. Our attempts to replicate their work (an official code is not available) have been fruitless.}, we created a new benchmark where we train a fully supervised generator using only longitudinal ADNI data. The results are shown in Table~\ref{tab: longitudinal benchmark}. We see that our method has slightly better performance than the longitudinal model. This is because the proposed model is trained on 786 subjects (cross-sectional data), while the longitudinal model is trained on a longitudinal cohort of ADNI of 98 subjects. This illustrates the benefit of using cross-sectional data. Note that our SSIM results are similar to those presented in \cite{ravi2019degenerative}.

\begin{table}[t]
\centering
\caption{Quantitative results of different choices of the $v_2$ dimension.}
\label{tab: ablation study of latent space}

\begin{tabular}{l|lllll}
\hline
                     & SSIM & PSNR & MSE  &PAD\\ \hline
$65 \times 1$            &   $0.73_{\pm 0.09}$   &  $23.6_{\pm 3.1}$    &  $0.065_{\pm 0.012}$ &$5.6_{\pm 4.1 }$        \\
$260 \times 1$ &   $0.76_{\pm 0.10}$   &    $24.9_{\pm 2.9}$  & $0.055_{\pm 0.012}$   & $5.3_{\pm 3.8}$      \\
\hline
$130 \times 1$ (ours) & $\mathbf{0.79^{*}_{\pm 0.08}}$    & $\mathbf{26.1^{*}_{\pm 2.6}}$     &   $\mathbf{0.042^{*}_{\pm 0.006}}$  &  $\mathbf{5.0_{\pm 3.6}}$        \\
 \hline

\end{tabular}
\end{table}

\subsubsection{Data augmentation for AD classification}
We explore whether we can use our model to generate synthetic data used to augment training sets for training an Alzheimer's disease classifier. We select 200 ADNI subjects as training data (100 AD, 100 CN), 40 subjects as validation data (20 AD, 20 CN), and 80 (40 AD, 40 CN) subjects as testing data. For each subject, there are 60 2D slices. Next, we train classifiers of the same VGG architecture to classify AD and CN subjects varying the composition of the training data combining real and synthetic data obtained by our generator. We always evaluate the classifiers on the same testing set. The synthetic data are generated from the training set using our proposed method by randomly selecting target ages larger than the original ages. As shown in Table~\ref{tab: AD classification accuacy}, we first train classifiers only on real data varying the size of the training data (1st and 2nd rows). Then we compose mixed sets of the same size of 200 subjects varying the ratio of real vs.\ synthetic data (3rd and 4th rows), e.g.\ 10\%+90\% means this set is composed of 10\% real data and 90\% synthetic data. Note here the 90\% synthetic data are not generated from the whole training set, but from the 10\% real data.


We can observe that when training on 10\% of real training data, the accuracy reduces by almost 40\% compared to when using the full training data. However, the performances improve when synthetic data are involved. The results demonstrate that our method can be used as data augmentation to improve AD classification especially when the training data are not sufficient.

Furthermore, we perform another experiment to demonstrate our model’s potential to improve the classification accuracy for specific age groups and thus target augmentation to treat data imbalance. We evaluate the classification model trained with 100\% real data on test set subjects of age 65 to 70 years old. We find an accuracy of 67.2\%, which is much lower than the overall accuracy (89.5\%, Table~\ref{tab: AD classification accuacy}). This may be likely due to training data imbalance: we have only 5 training subjects with age between 65 and 70 yrs. To alleviate this data imbalance, we use our model to generate 25 synthetic subjects with target ages between 65 and 70 yrs from younger subjects in the training set. Then we train a new AD classifier on 100\% real data and the 25 synthetic subjects, and evaluate its performance on the same testing and age group. Accuracy now increases to 80.1\% a substantial change from 67.2\%. 

\section{Conclusion} 
\label{sec: conclusion}
We present a method that learns to simulate subject-specific  aged images \textit{without} longitudinal data. It relies on a Generator to generate the images and a Discriminator that captures the joint distribution of brain images and clinical variables, i.e.\ age and health state (AD status). We propose an embedding mechanism to encode the information of age and health state into our network, and age-modulated and self-reconstruction losses to preserve \textit{subject identity}. We present qualitative results showing that our method is able to generate consistent and realistic images conditioned on the target age and health state. We evaluate with longitudinal data from ADNI for image quality and \textit{age accuracy}. We demonstrate on ADNI and Cam-CAN datasets that our model outperforms benchmarks both qualitatively and quantitatively and, via a series of ablations, illustrate the importance of each design decision.
\begin{table}[t]
\centering
\caption{Quantitative results of a longitudinal benchmark and our method.}
\label{tab: longitudinal benchmark}
\begin{tabular}{r|ccc}
\hline
\multicolumn{1}{l|}{} & SSIM      & PSNR     & MSE         \\ \hline
Longitudinal          & $0.72_{\pm 0.09}$ & $24.2_{\pm 3.0}$ & $0.076_{\pm 0.013}$ \\ \hline
Ours                  & $0.79_{\pm 0.08}$ & $26.1_{\pm 2.6}$ & $0.042_{\pm 0.006}$ \\ \hline
\end{tabular}
\end{table}
\begin{table*}[tb]
\centering
\caption{Quantitative results of VGG-based AD/CN classifiers trained on different datasets. The first two rows show results when trained on varying size of real training data, e.g. 10\% means this model is trained on 10\% of the real training data; the last two rows show results when trained on mixed datasets with different ratios of real and synthetic data, e.g. 10\%+90\% means this model is trained on 10\% real training data and 90\% synthetic data.}
\label{tab: AD classification accuacy}
\begin{tabular}{r|ccccc}
\hline
Real data                  & 10\%      & 30\%      & 50\%      & 70\%                           & 100\%                 \\ \hline
Accuracy (\%)                 & 51.3      & 55.7      & 64.6      & \multicolumn{1}{c|}{74.0}      & \multirow{3}{*}{89.5} \\ \cline{1-5}
Real data + synthetic data & 10\%+90\% & 30\%+70\% & 50\%+50\% & \multicolumn{1}{c|}{70\%+30\%} &                       \\ \cline{1-5}
Accuracy (\%)                 & 58.7      & 64.0      & 72.6      & \multicolumn{1}{c|}{81.7}      &                       \\ \hline
\end{tabular}
\end{table*}

\textbf{Potential applications.}
The proposed method has several potential applications. For example, a common problem in longitudinal studies is missing data due to patient dropout or poor-quality scans. The proposed method offers an opportunity to impute missing data at any time point. Furthermore, when there is insufficient longitudinal training data, the proposed method can be used to include cross-sectional data within a study. The simple experiment in \ref{sec:mtg} shows a glimpse of this potential.

This in turn will make further clinical analysis of ageing patterns, e.g.\ to evaluate the incidence of white matter hyperintensities~\citep{wardlaw2015white}, and large studies into neurodegenerative diseases, possible. Finally, from an AI perspective we advocated earlier on the paper about the importance of capturing and understanding current state from a machine learning perspective. In fact, recently this has been cast in a causal inference and counterfactual setting \cite{pawlowski2020deep}. While our work didn't explicitly use a causal inference framework, our generated outputs can be seen as counterfactuals. 


\textbf{Limitations and avenues for improvement.}
The notion of subject identity is context specific and we do note that others in the literature also follow the same simple assumptions we make. We do agree though that identity should be defined as what remains invariant under ageing and neurodegenerative disease.
Although we used several losses to help preserve subject identity of synthetic aged images, there is no guarantee that a subject's identity will be preserved, and new losses or mechanisms that could further improve identity preservation will be of high value. Unfortunately, without access to large data where we exhaustively explore all possible combinations of variables that we want to be equivalent  (to identity) or invariant (to age, pathology) preservation of identity can only be proxied.
The proposed model only considers predicting older brain images from young ones. However, performing both brain ageing and rejuvenation will provide more utility but will require more advanced design of encoding and network architecture.
The proposed method allows for change of health status between input and output images. However, it does not model change of health state in between input and output. This is a common limitation of current works in this area~\citep{ravi2019degenerative,pawlowski2020deep,rachmadi2019predicting}.
A potential solution is recursive image synthesis: generating a suitable intermediate image before generating the desired target output of an older age and state. Advances in architectures that improve image quality will enable such recursive image generation in the future.
Conditioning mechanisms that reliably embed prior information into neural networks enabling finer control over the outputs of models are of considerable interest in deep learning. In this paper we design a simple yet effective way to encode both \textit{age} (continuous) and \textit{AD status} (ordinal) factors into the image generation network. However, as classification of MCI is challenging, use of further (fine-grained) clinical information (e.g.\ clinical score) to reflect health status can be of benefit. Incorporating additional clinical variables, e.g.\ gender, genotypes, etc., can become inefficient with our current approach as it may involve more dense layers. While new techniques are available~\citep{huang2017arbitrary,perez2018film,park2019semantic,lee2019tetris} and some prior examples on few conditioning variables~\citep{jacenkow2019conditioning} or disentanglement~\citep{chartsias2019disentangled} are promising, their utility in integrating clinical variables, and replacing the need for ordinal pre-encoding of continuous or ordinal variables, with imaging data is under investigation.  Although we used brain data, the approach could be extended to other organs.
Furthermore, here we focus on the use of cross-sectional data to train a model to predict aged brain images. If longitudinal data are also available, e.g.\ within a large study aggregating several data sources, model performance could be further improved by introducing supervised losses; however, adding more losses requires that they are well balanced --a known problem in semi-supervised learning~\citep{sener2018multi}. We showed that our synthetic volumes (composed by stacking 2D images) can achieve good quality. In all our attempts with 3D architectures, the parameter space exploded due to their size. We expect that advances in network compression~\citep{han2015deep} will eventually permit us to adapt a 3D design which should further improve visual quality and consistency of the approach and allow us to repeat our atrophy analysis not only in the MTG.

\section{Acknowledgements}
This work was supported by the University of Edinburgh by PhD studentships to T. Xia and A. Chartsias. This work was supported by The Alan Turing Institute under the EPSRC grant EP/N510129/1. We thank Nvidia for donating a Titan-X GPU. S.A. Tsaftaris acknowledges the support of Canon Medical and the Royal Academy of Engineering and the Research Chairs and Senior Research Fellowships scheme (grant RCSRF1819/8/25). Data collection and sharing for this project was funded by the Alzheimer's Disease Neuroimaging Initiative (ADNI) (NIH grant U01 AG024904) and Department of Defense ADNI (award number W81XWH-12-2-0012). ADNI is funded by the National Institute on Aging, the National Institute of Biomedical Imaging and Bioengineering.  Private sector contributions are facilitated by the Foundation for the National Institutes of Health (www.fnih.org).  ADNI data are disseminated by the Laboratory for Neuro Imaging at the University of Southern California.

\bibliographystyle{model2-names.bst}\biboptions{authoryear}
\bibliography{reference}

\begin{thebibliography}{72}
\expandafter\ifx\csname natexlab\endcsname\relax\def\natexlab#1{#1}\fi
\providecommand{\url}[1]{\texttt{#1}}
\providecommand{\href}[2]{#2}
\providecommand{\path}[1]{#1}
\providecommand{\DOIprefix}{doi:}
\providecommand{\ArXivprefix}{arXiv:}
\providecommand{\URLprefix}{URL: }
\providecommand{\Pubmedprefix}{pmid:}
\providecommand{\doi}[1]{\href{http://dx.doi.org/#1}{\path{#1}}}
\providecommand{\Pubmed}[1]{\href{pmid:#1}{\path{#1}}}
\providecommand{\bibinfo}[2]{#2}
\ifx\xfnm\relax \def\xfnm[#1]{\unskip,\space#1}\fi
\bibitem[{Arjovsky et~al.(2017)Arjovsky, Chintala and
  Bottou}]{arjovsky2017wasserstein}
\bibinfo{author}{Arjovsky, M.}, \bibinfo{author}{Chintala, S.},
  \bibinfo{author}{Bottou, L.}, \bibinfo{year}{2017}.
\newblock \bibinfo{title}{{Wasserstein generative adversarial networks}}, in:
  \bibinfo{booktitle}{ICML}.
\bibitem[{Avants et~al.(2008)Avants, Epstein, Grossman and
  Gee}]{avants2008symmetric}
\bibinfo{author}{Avants, B.B.}, \bibinfo{author}{Epstein, C.L.},
  \bibinfo{author}{Grossman, M.}, \bibinfo{author}{Gee, J.C.},
  \bibinfo{year}{2008}.
\newblock \bibinfo{title}{Symmetric diffeomorphic image registration with
  cross-correlation: evaluating automated labeling of elderly and
  neurodegenerative brain}.
\newblock \bibinfo{journal}{Medical image analysis} \bibinfo{volume}{12},
  \bibinfo{pages}{26--41}.
\bibitem[{Baumgartner et~al.(2018)Baumgartner, Koch, Can~Tezcan, Xi~Ang and
  Konukoglu}]{baumgartner2018visual}
\bibinfo{author}{Baumgartner, C.F.}, \bibinfo{author}{Koch, L.M.},
  \bibinfo{author}{Can~Tezcan, K.}, \bibinfo{author}{Xi~Ang, J.},
  \bibinfo{author}{Konukoglu, E.}, \bibinfo{year}{2018}.
\newblock \bibinfo{title}{Visual feature attribution using wasserstein {GANs}},
  in: \bibinfo{booktitle}{CVPR}, pp. \bibinfo{pages}{8309--19}.
\bibitem[{Bowles et~al.(2018)Bowles, Gunn, Hammers and
  Rueckert}]{bowles2018modelling}
\bibinfo{author}{Bowles, C.}, \bibinfo{author}{Gunn, R.},
  \bibinfo{author}{Hammers, A.}, \bibinfo{author}{Rueckert, D.},
  \bibinfo{year}{2018}.
\newblock \bibinfo{title}{{Modelling the progression of Alzheimer's disease in
  MRI using generative adversarial networks}}, in: \bibinfo{booktitle}{Medical
  Imaging 2018: Image Processing}.
\bibitem[{Camara et~al.(2006)Camara, Schweiger, Scahill, Crum, Sneller,
  Schnabel, Ridgway, Cash, Hill and Fox}]{camara2006phenomenological}
\bibinfo{author}{Camara, O.}, \bibinfo{author}{Schweiger, M.},
  \bibinfo{author}{Scahill, R.I.}, \bibinfo{author}{Crum, W.R.},
  \bibinfo{author}{Sneller, B.I.}, \bibinfo{author}{Schnabel, J.A.},
  \bibinfo{author}{Ridgway, G.R.}, \bibinfo{author}{Cash, D.M.},
  \bibinfo{author}{Hill, D.L.}, \bibinfo{author}{Fox, N.C.},
  \bibinfo{year}{2006}.
\newblock \bibinfo{title}{Phenomenological model of diffuse global and regional
  atrophy using finite-element methods}.
\newblock \bibinfo{journal}{TMI} \bibinfo{volume}{25},
  \bibinfo{pages}{1417--1430}.
\bibitem[{Chartsias et~al.(2019)Chartsias, Joyce, Papanastasiou, Semple,
  Williams, Newby, Dharmakumar and Tsaftaris}]{chartsias2019disentangled}
\bibinfo{author}{Chartsias, A.}, \bibinfo{author}{Joyce, T.},
  \bibinfo{author}{Papanastasiou, G.}, \bibinfo{author}{Semple, S.},
  \bibinfo{author}{Williams, M.}, \bibinfo{author}{Newby, D.E.},
  \bibinfo{author}{Dharmakumar, R.}, \bibinfo{author}{Tsaftaris, S.A.},
  \bibinfo{year}{2019}.
\newblock \bibinfo{title}{Disentangled representation learning in cardiac image
  analysis}.
\newblock \bibinfo{journal}{Medical Image Analysis} \bibinfo{volume}{58},
  \bibinfo{pages}{101535}.
\bibitem[{Chollet et~al.(2015)}]{chollet2015keras}
\bibinfo{author}{Chollet, F.}, et~al., \bibinfo{year}{2015}.
\newblock \bibinfo{title}{Keras}.
\newblock \bibinfo{howpublished}{\url{https://keras.io}}.
\bibitem[{Cole and Franke(2017)}]{cole2017predicting}
\bibinfo{author}{Cole, J.H.}, \bibinfo{author}{Franke, K.},
  \bibinfo{year}{2017}.
\newblock \bibinfo{title}{Predicting age using neuroimaging: innovative brain
  ageing biomarkers}.
\newblock \bibinfo{journal}{{Trends in Neurosciences}} \bibinfo{volume}{40},
  \bibinfo{pages}{681--690}.
\bibitem[{Cole et~al.(2015)Cole, Leech, Sharp and
  Initiative}]{cole2015prediction}
\bibinfo{author}{Cole, J.H.}, \bibinfo{author}{Leech, R.},
  \bibinfo{author}{Sharp, D.J.}, \bibinfo{author}{Initiative, A.D.N.},
  \bibinfo{year}{2015}.
\newblock \bibinfo{title}{Prediction of brain age suggests accelerated atrophy
  after traumatic brain injury}.
\newblock \bibinfo{journal}{Annals of neurology} \bibinfo{volume}{77},
  \bibinfo{pages}{571--581}.
\bibitem[{Cole et~al.(2019)Cole, Marioni, Harris and Deary}]{cole2019brain}
\bibinfo{author}{Cole, J.H.}, \bibinfo{author}{Marioni, R.E.},
  \bibinfo{author}{Harris, S.E.}, \bibinfo{author}{Deary, I.J.},
  \bibinfo{year}{2019}.
\newblock \bibinfo{title}{Brain age and other bodily ‘ages’: implications
  for neuropsychiatry}.
\newblock \bibinfo{journal}{{Molecular Psychiatry}} \bibinfo{volume}{24},
  \bibinfo{pages}{266}.
\bibitem[{Cole et~al.(2017)Cole, Ritchie, Bastin, Hern{\'a}ndez, Maniega,
  Royle, Corley, Pattie, Harris, Zhang et~al.}]{cole2017mortality}
\bibinfo{author}{Cole, J.H.}, \bibinfo{author}{Ritchie, S.J.},
  \bibinfo{author}{Bastin, M.E.}, \bibinfo{author}{Hern{\'a}ndez, M.V.},
  \bibinfo{author}{Maniega, S.M.}, \bibinfo{author}{Royle, N.},
  \bibinfo{author}{Corley, J.}, \bibinfo{author}{Pattie, A.},
  \bibinfo{author}{Harris, S.E.}, \bibinfo{author}{Zhang, Q.}, et~al.,
  \bibinfo{year}{2017}.
\newblock \bibinfo{title}{Brain age predicts mortality}.
\newblock \bibinfo{journal}{{Molecular Psychiatry}} .
\bibitem[{Coleman~Jr et~al.(2014)Coleman~Jr, Liu, Oguz, Styner and
  Crews}]{coleman2014adolescent}
\bibinfo{author}{Coleman~Jr, L.G.}, \bibinfo{author}{Liu, W.},
  \bibinfo{author}{Oguz, I.}, \bibinfo{author}{Styner, M.},
  \bibinfo{author}{Crews, F.T.}, \bibinfo{year}{2014}.
\newblock \bibinfo{title}{Adolescent binge ethanol treatment alters adult brain
  regional volumes, cortical extracellular matrix protein and behavioral
  flexibility}.
\newblock \bibinfo{journal}{{Pharmacology Biochemistry and Behavior}}
  \bibinfo{volume}{116}, \bibinfo{pages}{142--151}.
\bibitem[{Costafreda et~al.(2011)Costafreda, Dinov, Tu, Shi, Liu, Kloszewska,
  Mecocci, Soininen, Tsolaki, Vellas et~al.}]{costafreda2011automated}
\bibinfo{author}{Costafreda, S.G.}, \bibinfo{author}{Dinov, I.D.},
  \bibinfo{author}{Tu, Z.}, \bibinfo{author}{Shi, Y.}, \bibinfo{author}{Liu,
  C.Y.}, \bibinfo{author}{Kloszewska, I.}, \bibinfo{author}{Mecocci, P.},
  \bibinfo{author}{Soininen, H.}, \bibinfo{author}{Tsolaki, M.},
  \bibinfo{author}{Vellas, B.}, et~al., \bibinfo{year}{2011}.
\newblock \bibinfo{title}{Automated hippocampal shape analysis predicts the
  onset of dementia in mild cognitive impairment}.
\newblock \bibinfo{journal}{Neuroimage} \bibinfo{volume}{56},
  \bibinfo{pages}{212--219}.
\bibitem[{Davis et~al.(2010)Davis, Fletcher, Bullitt and
  Joshi}]{davis2010population}
\bibinfo{author}{Davis, B.C.}, \bibinfo{author}{Fletcher, P.T.},
  \bibinfo{author}{Bullitt, E.}, \bibinfo{author}{Joshi, S.},
  \bibinfo{year}{2010}.
\newblock \bibinfo{title}{Population shape regression from random design data}.
\newblock \bibinfo{journal}{IJCV} .
\bibitem[{Fan et~al.(2016)Fan, Li, Zhuo, Zhang, Wang, Chen, Yang, Chu, Xie,
  Laird et~al.}]{fan2016human}
\bibinfo{author}{Fan, L.}, \bibinfo{author}{Li, H.}, \bibinfo{author}{Zhuo,
  J.}, \bibinfo{author}{Zhang, Y.}, \bibinfo{author}{Wang, J.},
  \bibinfo{author}{Chen, L.}, \bibinfo{author}{Yang, Z.}, \bibinfo{author}{Chu,
  C.}, \bibinfo{author}{Xie, S.}, \bibinfo{author}{Laird, A.R.}, et~al.,
  \bibinfo{year}{2016}.
\newblock \bibinfo{title}{The human brainnetome atlas: a new brain atlas based
  on connectional architecture}.
\newblock \bibinfo{journal}{Cerebral cortex} \bibinfo{volume}{26},
  \bibinfo{pages}{3508--3526}.
\bibitem[{Farokhian et~al.(2017)Farokhian, Yang, Beheshti, Matsuda and
  Wu}]{farokhian2017age}
\bibinfo{author}{Farokhian, F.}, \bibinfo{author}{Yang, C.},
  \bibinfo{author}{Beheshti, I.}, \bibinfo{author}{Matsuda, H.},
  \bibinfo{author}{Wu, S.}, \bibinfo{year}{2017}.
\newblock \bibinfo{title}{Age-related gray and white matter changes in normal
  adult brains}.
\newblock \bibinfo{journal}{Aging and disease} \bibinfo{volume}{8},
  \bibinfo{pages}{899}.
\bibitem[{Fjell and Walhovd(2010)}]{fjell2010structural}
\bibinfo{author}{Fjell, A.M.}, \bibinfo{author}{Walhovd, K.B.},
  \bibinfo{year}{2010}.
\newblock \bibinfo{title}{Structural brain changes in aging: courses, causes
  and cognitive consequences}.
\newblock \bibinfo{journal}{Reviews in the Neurosciences} \bibinfo{volume}{21},
  \bibinfo{pages}{187--222}.
\bibitem[{{Franke, Katja and Ziegler, Gabriel and Kl{\"o}ppel, Stefan and
  Gaser, Christian and Alzheimer's Disease Neuroimaging Initiative and
  others}(2010)}]{franke2010estimating}
\bibinfo{author}{{Franke, Katja and Ziegler, Gabriel and Kl{\"o}ppel, Stefan
  and Gaser, Christian and Alzheimer's Disease Neuroimaging Initiative and
  others}}, \bibinfo{year}{2010}.
\newblock \bibinfo{title}{{Estimating the age of healthy subjects from
  {T1-weighted MRI} scans using kernel methods: exploring the influence of
  various parameters}}.
\newblock \bibinfo{journal}{Neuroimage} \bibinfo{volume}{50},
  \bibinfo{pages}{883--892}.
\bibitem[{Good et~al.(2001)Good, Johnsrude, Ashburner, Henson, Friston and
  Frackowiak}]{good2001voxel}
\bibinfo{author}{Good, C.D.}, \bibinfo{author}{Johnsrude, I.S.},
  \bibinfo{author}{Ashburner, J.}, \bibinfo{author}{Henson, R.N.},
  \bibinfo{author}{Friston, K.J.}, \bibinfo{author}{Frackowiak, R.S.},
  \bibinfo{year}{2001}.
\newblock \bibinfo{title}{{A voxel-based morphometric study of ageing in 465
  normal adult human brains}}.
\newblock \bibinfo{journal}{Neuroimage} \bibinfo{volume}{14},
  \bibinfo{pages}{21--36}.
\bibitem[{Goodfellow et~al.(2014)Goodfellow, Pouget-Abadie, Mirza, Xu,
  Warde-Farley, Ozair, Courville and Bengio}]{goodfellow2014generative}
\bibinfo{author}{Goodfellow, I.}, \bibinfo{author}{Pouget-Abadie, J.},
  \bibinfo{author}{Mirza, M.}, \bibinfo{author}{Xu, B.},
  \bibinfo{author}{Warde-Farley, D.}, \bibinfo{author}{Ozair, S.},
  \bibinfo{author}{Courville, A.}, \bibinfo{author}{Bengio, Y.},
  \bibinfo{year}{2014}.
\newblock \bibinfo{title}{Generative adversarial nets}, in:
  \bibinfo{booktitle}{NeurIPS}, pp. \bibinfo{pages}{2672--2680}.
\bibitem[{Gulrajani et~al.(2017)Gulrajani, Ahmed, Arjovsky, Dumoulin and
  Courville}]{gulrajani2017improved}
\bibinfo{author}{Gulrajani, I.}, \bibinfo{author}{Ahmed, F.},
  \bibinfo{author}{Arjovsky, M.}, \bibinfo{author}{Dumoulin, V.},
  \bibinfo{author}{Courville, A.C.}, \bibinfo{year}{2017}.
\newblock \bibinfo{title}{{Improved training of Wasserstein GANs}}, in:
  \bibinfo{booktitle}{NeurIPS}, pp. \bibinfo{pages}{5767--5777}.
\bibitem[{Guo et~al.(2014)Guo, Zhang, Zhou, Wang, Yao, Yuan, An, Dai, Wang,
  Zhang et~al.}]{guo2014grey}
\bibinfo{author}{Guo, Y.}, \bibinfo{author}{Zhang, Z.}, \bibinfo{author}{Zhou,
  B.}, \bibinfo{author}{Wang, P.}, \bibinfo{author}{Yao, H.},
  \bibinfo{author}{Yuan, M.}, \bibinfo{author}{An, N.}, \bibinfo{author}{Dai,
  H.}, \bibinfo{author}{Wang, L.}, \bibinfo{author}{Zhang, X.}, et~al.,
  \bibinfo{year}{2014}.
\newblock \bibinfo{title}{Grey-matter volume as a potential feature for the
  classification of alzheimer’s disease and mild cognitive impairment: an
  exploratory study}.
\newblock \bibinfo{journal}{Neuroscience bulletin} \bibinfo{volume}{30},
  \bibinfo{pages}{477--489}.
\bibitem[{Han et~al.(2016)Han, Mao and Dally}]{han2015deep}
\bibinfo{author}{Han, S.}, \bibinfo{author}{Mao, H.}, \bibinfo{author}{Dally,
  W.J.}, \bibinfo{year}{2016}.
\newblock \bibinfo{title}{Deep compression: Compressing deep neural networks
  with pruning, trained quantization and huffman coding}, in:
  \bibinfo{booktitle}{International Conference on Learning Representations
  (ICLR)}.
\bibitem[{He et~al.(2016)He, Zhang, Ren and Sun}]{he2016deep}
\bibinfo{author}{He, K.}, \bibinfo{author}{Zhang, X.}, \bibinfo{author}{Ren,
  S.}, \bibinfo{author}{Sun, J.}, \bibinfo{year}{2016}.
\newblock \bibinfo{title}{{Deep residual learning for image recognition}}, in:
  \bibinfo{booktitle}{CVPR}.
\bibitem[{Huang and Belongie(2017)}]{huang2017arbitrary}
\bibinfo{author}{Huang, X.}, \bibinfo{author}{Belongie, S.},
  \bibinfo{year}{2017}.
\newblock \bibinfo{title}{Arbitrary style transfer in real-time with adaptive
  instance normalization}, in: \bibinfo{booktitle}{CVPR}, pp.
  \bibinfo{pages}{1501--1510}.
\bibitem[{Huizinga et~al.(2018)Huizinga, Poot, Vernooij, Roshchupkin, Bron,
  Ikram, Rueckert, Niessen, Klein, Initiative et~al.}]{huizinga2018spatio}
\bibinfo{author}{Huizinga, W.}, \bibinfo{author}{Poot, D.H.},
  \bibinfo{author}{Vernooij, M.W.}, \bibinfo{author}{Roshchupkin, G.},
  \bibinfo{author}{Bron, E.}, \bibinfo{author}{Ikram, M.A.},
  \bibinfo{author}{Rueckert, D.}, \bibinfo{author}{Niessen, W.J.},
  \bibinfo{author}{Klein, S.}, \bibinfo{author}{Initiative, A.D.N.}, et~al.,
  \bibinfo{year}{2018}.
\newblock \bibinfo{title}{{A spatio-temporal reference model of the aging
  brain}}.
\newblock \bibinfo{journal}{NeuroImage} \bibinfo{volume}{169},
  \bibinfo{pages}{11--22}.
\bibitem[{Jacenkow et~al.(2019)Jacenkow, Chartsias, Mohr and
  Tsaftaris}]{jacenkow2019conditioning}
\bibinfo{author}{Jacenkow, G.}, \bibinfo{author}{Chartsias, A.},
  \bibinfo{author}{Mohr, B.}, \bibinfo{author}{Tsaftaris, S.A.},
  \bibinfo{year}{2019}.
\newblock \bibinfo{title}{Conditioning convolutional segmentation architectures
  with non-imaging data}, in: \bibinfo{booktitle}{MIDL}.
\bibitem[{Jack et~al.(1998)Jack, Petersen, Xu, O'Brien, Smith, Ivnik, Tangalos
  and Kokmen}]{jack1998rate}
\bibinfo{author}{Jack, C.}, \bibinfo{author}{Petersen, R.C.},
  \bibinfo{author}{Xu, Y.}, \bibinfo{author}{O'Brien, P.C.},
  \bibinfo{author}{Smith, G.E.}, \bibinfo{author}{Ivnik, R.J.},
  \bibinfo{author}{Tangalos, E.G.}, \bibinfo{author}{Kokmen, E.},
  \bibinfo{year}{1998}.
\newblock \bibinfo{title}{{Rate of medial temporal lobe atrophy in typical
  aging and Alzheimer's disease}}.
\newblock \bibinfo{journal}{Neurology} \bibinfo{volume}{51},
  \bibinfo{pages}{993--999}.
\bibitem[{Jonsson et~al.(2019)Jonsson, Bjornsdottir, Thorgeirsson, Ellingsen,
  Walters, Gudbjartsson, Stefansson, Stefansson and
  Ulfarsson}]{jonsson2019deep}
\bibinfo{author}{Jonsson, B.}, \bibinfo{author}{Bjornsdottir, G.},
  \bibinfo{author}{Thorgeirsson, T.}, \bibinfo{author}{Ellingsen, L.},
  \bibinfo{author}{Walters, G.B.}, \bibinfo{author}{Gudbjartsson, D.},
  \bibinfo{author}{Stefansson, H.}, \bibinfo{author}{Stefansson, K.},
  \bibinfo{author}{Ulfarsson, M.}, \bibinfo{year}{2019}.
\newblock \bibinfo{title}{Deep learning based brain age prediction uncovers
  associated sequence variants}.
\newblock \bibinfo{journal}{bioRxiv} , \bibinfo{pages}{595801}.
\bibitem[{Khanal et~al.(2017)Khanal, Ayache and Pennec}]{khanal2017simulating}
\bibinfo{author}{Khanal, B.}, \bibinfo{author}{Ayache, N.},
  \bibinfo{author}{Pennec, X.}, \bibinfo{year}{2017}.
\newblock \bibinfo{title}{{Simulating longitudinal brain MRIs with known volume
  changes and realistic variations in image intensity}}.
\newblock \bibinfo{journal}{{Frontiers in Neuroscience}} \bibinfo{volume}{11},
  \bibinfo{pages}{132}.
\bibitem[{Kingma and Ba(2015)}]{kingma2015adam}
\bibinfo{author}{Kingma, D.P.}, \bibinfo{author}{Ba, J.}, \bibinfo{year}{2015}.
\newblock \bibinfo{title}{Adam: A method for stochastic optimization}.
\newblock \bibinfo{journal}{International Conference on Learning
  Representations} .
\bibitem[{Lee et~al.(2019)Lee, Petersen, Pawlowski, Glocker and
  Schaap}]{lee2019tetris}
\bibinfo{author}{Lee, M.C.H.}, \bibinfo{author}{Petersen, K.},
  \bibinfo{author}{Pawlowski, N.}, \bibinfo{author}{Glocker, B.},
  \bibinfo{author}{Schaap, M.}, \bibinfo{year}{2019}.
\newblock \bibinfo{title}{{Tetris: Template transformer networks for image
  segmentation with shape priors}}.
\newblock \bibinfo{journal}{TMI} \bibinfo{volume}{38},
  \bibinfo{pages}{2596--2606}.
\bibitem[{L{\'o}pez-Ot{\'\i}n et~al.(2013)L{\'o}pez-Ot{\'\i}n, Blasco,
  Partridge, Serrano and Kroemer}]{lopez2013hallmarks}
\bibinfo{author}{L{\'o}pez-Ot{\'\i}n, C.}, \bibinfo{author}{Blasco, M.A.},
  \bibinfo{author}{Partridge, L.}, \bibinfo{author}{Serrano, M.},
  \bibinfo{author}{Kroemer, G.}, \bibinfo{year}{2013}.
\newblock \bibinfo{title}{The hallmarks of aging}.
\newblock \bibinfo{journal}{Cell} \bibinfo{volume}{153},
  \bibinfo{pages}{1194--1217}.
\bibitem[{Lorenzi et~al.(2015)Lorenzi, Pennec, Frisoni, Ayache, Initiative
  et~al.}]{lorenzi2015disentangling}
\bibinfo{author}{Lorenzi, M.}, \bibinfo{author}{Pennec, X.},
  \bibinfo{author}{Frisoni, G.B.}, \bibinfo{author}{Ayache, N.},
  \bibinfo{author}{Initiative, A.D.N.}, et~al., \bibinfo{year}{2015}.
\newblock \bibinfo{title}{Disentangling normal aging from alzheimer's disease
  in structural magnetic resonance images}.
\newblock \bibinfo{journal}{Neurobiology of aging} \bibinfo{volume}{36},
  \bibinfo{pages}{S42--S52}.
\bibitem[{Mattson and Arumugam(2018)}]{mattson2018hallmarks}
\bibinfo{author}{Mattson, M.P.}, \bibinfo{author}{Arumugam, T.V.},
  \bibinfo{year}{2018}.
\newblock \bibinfo{title}{Hallmarks of brain aging: adaptive and pathological
  modification by metabolic states}.
\newblock \bibinfo{journal}{{Cell Metabolism}} \bibinfo{volume}{27},
  \bibinfo{pages}{1176--1199}.
\bibitem[{Mietchen and Gaser(2009)}]{mietchen2009computational}
\bibinfo{author}{Mietchen, D.}, \bibinfo{author}{Gaser, C.},
  \bibinfo{year}{2009}.
\newblock \bibinfo{title}{Computational morphometry for detecting changes in
  brain structure due to development, aging, learning, disease and evolution}.
\newblock \bibinfo{journal}{{Frontiers in Neuroinformatics}}
  \bibinfo{volume}{3}, \bibinfo{pages}{25}.
\bibitem[{Milana(2017)}]{milana2017deep}
\bibinfo{author}{Milana, D.}, \bibinfo{year}{2017}.
\newblock \bibinfo{title}{{Deep generative models for predicting Alzheimer's
  disease progression from MR data}}.
\newblock Master's thesis. Politecnico Di Milano.
\bibitem[{Mirza and Osindero(2014)}]{mirza2014conditional}
\bibinfo{author}{Mirza, M.}, \bibinfo{author}{Osindero, S.},
  \bibinfo{year}{2014}.
\newblock \bibinfo{title}{{Conditional generative adversarial nets}}.
\newblock \bibinfo{journal}{arXiv preprint arXiv:1411.1784} .
\bibitem[{Modat et~al.(2014)Modat, Simpson, Cardoso, Cash, Toussaint, Fox and
  Ourselin}]{modat2014simulating}
\bibinfo{author}{Modat, M.}, \bibinfo{author}{Simpson, I.J.},
  \bibinfo{author}{Cardoso, M.J.}, \bibinfo{author}{Cash, D.M.},
  \bibinfo{author}{Toussaint, N.}, \bibinfo{author}{Fox, N.C.},
  \bibinfo{author}{Ourselin, S.}, \bibinfo{year}{2014}.
\newblock \bibinfo{title}{{Simulating neurodegeneration through longitudinal
  population analysis of structural and diffusion weighted MRI data}}, in:
  \bibinfo{booktitle}{MICCAI}, \bibinfo{organization}{Springer}. pp.
  \bibinfo{pages}{57--64}.
\bibitem[{Oramas et~al.(2018)Oramas, Wang and Tuytelaars}]{oramas2018visual}
\bibinfo{author}{Oramas, J.}, \bibinfo{author}{Wang, K.},
  \bibinfo{author}{Tuytelaars, T.}, \bibinfo{year}{2018}.
\newblock \bibinfo{title}{Visual explanation by interpretation: Improving
  visual feedback capabilities of deep neural networks}, in:
  \bibinfo{booktitle}{International Conference on Learning Representations}.
\bibitem[{Park et~al.(2019)Park, Liu, Wang and Zhu}]{park2019semantic}
\bibinfo{author}{Park, T.}, \bibinfo{author}{Liu, M.Y.}, \bibinfo{author}{Wang,
  T.C.}, \bibinfo{author}{Zhu, J.Y.}, \bibinfo{year}{2019}.
\newblock \bibinfo{title}{Semantic image synthesis with spatially-adaptive
  normalization}, in: \bibinfo{booktitle}{CVPR}, pp.
  \bibinfo{pages}{2337--2346}.
\bibitem[{Pawlowski et~al.(2020)Pawlowski, Coelho~de Castro and
  Glocker}]{pawlowski2020deep}
\bibinfo{author}{Pawlowski, N.}, \bibinfo{author}{Coelho~de Castro, D.},
  \bibinfo{author}{Glocker, B.}, \bibinfo{year}{2020}.
\newblock \bibinfo{title}{Deep structural causal models for tractable
  counterfactual inference}.
\newblock \bibinfo{journal}{Advances in Neural Information Processing Systems}
  \bibinfo{volume}{33}.
\bibitem[{Peng et~al.(2021)Peng, Gong, Beckmann, Vedaldi and
  Smith}]{peng2021accurate}
\bibinfo{author}{Peng, H.}, \bibinfo{author}{Gong, W.},
  \bibinfo{author}{Beckmann, C.F.}, \bibinfo{author}{Vedaldi, A.},
  \bibinfo{author}{Smith, S.M.}, \bibinfo{year}{2021}.
\newblock \bibinfo{title}{Accurate brain age prediction with lightweight deep
  neural networks}.
\newblock \bibinfo{journal}{Medical Image Analysis} \bibinfo{volume}{68},
  \bibinfo{pages}{101871}.
\bibitem[{Perez et~al.(2018)Perez, Strub, De~Vries, Dumoulin and
  Courville}]{perez2018film}
\bibinfo{author}{Perez, E.}, \bibinfo{author}{Strub, F.},
  \bibinfo{author}{De~Vries, H.}, \bibinfo{author}{Dumoulin, V.},
  \bibinfo{author}{Courville, A.}, \bibinfo{year}{2018}.
\newblock \bibinfo{title}{{FILM: Visual reasoning with a general conditioning
  layer}}, in: \bibinfo{booktitle}{AAAI}.
\bibitem[{Petersen et~al.(2010)Petersen, Aisen, Beckett, Donohue, Gamst,
  Harvey, Jack, Jagust, Shaw, Toga et~al.}]{petersen2010alzheimer}
\bibinfo{author}{Petersen, R.C.}, \bibinfo{author}{Aisen, P.},
  \bibinfo{author}{Beckett, L.A.}, \bibinfo{author}{Donohue, M.},
  \bibinfo{author}{Gamst, A.}, \bibinfo{author}{Harvey, D.J.},
  \bibinfo{author}{Jack, C.}, \bibinfo{author}{Jagust, W.},
  \bibinfo{author}{Shaw, L.}, \bibinfo{author}{Toga, A.}, et~al.,
  \bibinfo{year}{2010}.
\newblock \bibinfo{title}{{Alzheimer's disease neuroimaging initiative (ADNI):
  clinical characterization}}.
\newblock \bibinfo{journal}{Neurology} \bibinfo{volume}{74},
  \bibinfo{pages}{201--209}.
\bibitem[{Pieperhoff et~al.(2008)Pieperhoff, S{\"u}dmeyer, H{\"o}mke, Zilles,
  Schnitzler and Amunts}]{pieperhoff2008detection}
\bibinfo{author}{Pieperhoff, P.}, \bibinfo{author}{S{\"u}dmeyer, M.},
  \bibinfo{author}{H{\"o}mke, L.}, \bibinfo{author}{Zilles, K.},
  \bibinfo{author}{Schnitzler, A.}, \bibinfo{author}{Amunts, K.},
  \bibinfo{year}{2008}.
\newblock \bibinfo{title}{Detection of structural changes of the human brain in
  longitudinally acquired mr images by deformation field morphometry:
  methodological analysis, validation and application}.
\newblock \bibinfo{journal}{NeuroImage} \bibinfo{volume}{43},
  \bibinfo{pages}{269--287}.
\bibitem[{Rachmadi et~al.(2020)Rachmadi, Vald{\'e}s-Hern{\'a}ndez, Makin,
  Wardlaw and Komura}]{rachmadi2020automatic}
\bibinfo{author}{Rachmadi, M.F.}, \bibinfo{author}{Vald{\'e}s-Hern{\'a}ndez,
  M.d.C.}, \bibinfo{author}{Makin, S.}, \bibinfo{author}{Wardlaw, J.},
  \bibinfo{author}{Komura, T.}, \bibinfo{year}{2020}.
\newblock \bibinfo{title}{Automatic spatial estimation of white matter
  hyperintensities evolution in brain mri using disease evolution predictor
  deep neural networks}.
\newblock \bibinfo{journal}{Medical Image Analysis} , \bibinfo{pages}{101712}.
\bibitem[{Rachmadi et~al.(2019)Rachmadi, Vald{\'e}s-Hern{\'a}ndez, Makin,
  Wardlaw and Komura}]{rachmadi2019predicting}
\bibinfo{author}{Rachmadi, M.F.}, \bibinfo{author}{Vald{\'e}s-Hern{\'a}ndez,
  M.d.C.}, \bibinfo{author}{Makin, S.}, \bibinfo{author}{Wardlaw, J.M.},
  \bibinfo{author}{Komura, T.}, \bibinfo{year}{2019}.
\newblock \bibinfo{title}{{Predicting the Evolution of White Matter
  Hyperintensities in Brain MRI using Generative Adversarial Networks and
  Irregularity Map}}.
\newblock \bibinfo{journal}{MICCAI} .
\bibitem[{Ravi et~al.(2019a)Ravi, Alexander and Oxtoby}]{ravi2019degenerative}
\bibinfo{author}{Ravi, D.}, \bibinfo{author}{Alexander, D.C.},
  \bibinfo{author}{Oxtoby, N.P.}, \bibinfo{year}{2019}a.
\newblock \bibinfo{title}{{Degenerative Adversarial NeuroImage Nets: Generating
  Images that Mimic Disease Progression}}.
\newblock \bibinfo{journal}{MICCAI} .
\bibitem[{Ravi et~al.(2019b)Ravi, Blumberg, Mengoudi, Xu, Alexander and
  Oxtoby}]{ravi2019degenerative1}
\bibinfo{author}{Ravi, D.}, \bibinfo{author}{Blumberg, S.B.},
  \bibinfo{author}{Mengoudi, K.}, \bibinfo{author}{Xu, M.},
  \bibinfo{author}{Alexander, D.C.}, \bibinfo{author}{Oxtoby, N.P.},
  \bibinfo{year}{2019}b.
\newblock \bibinfo{title}{Degenerative adversarial neuroimage nets for 4d
  simulations: Application in longitudinal mri}.
\newblock \bibinfo{journal}{arXiv preprint arXiv:1912.01526} .
\bibitem[{Ronneberger et~al.(2015)Ronneberger, Fischer and
  Brox}]{ronneberger2015u}
\bibinfo{author}{Ronneberger, O.}, \bibinfo{author}{Fischer, P.},
  \bibinfo{author}{Brox, T.}, \bibinfo{year}{2015}.
\newblock \bibinfo{title}{U-net: Convolutional networks for biomedical image
  segmentation}, in: \bibinfo{booktitle}{MICCAI},
  \bibinfo{organization}{Springer}. pp. \bibinfo{pages}{234--241}.
\bibitem[{Sener and Koltun(2018)}]{sener2018multi}
\bibinfo{author}{Sener, O.}, \bibinfo{author}{Koltun, V.},
  \bibinfo{year}{2018}.
\newblock \bibinfo{title}{Multi-task learning as multi-objective optimization},
  in: \bibinfo{booktitle}{NeurIPS}, pp. \bibinfo{pages}{527--538}.
\bibitem[{Serag et~al.(2012)Serag, Aljabar, Ball, Counsell, Boardman,
  Rutherford, Edwards, Hajnal and Rueckert}]{serag2012construction}
\bibinfo{author}{Serag, A.}, \bibinfo{author}{Aljabar, P.},
  \bibinfo{author}{Ball, G.}, \bibinfo{author}{Counsell, S.J.},
  \bibinfo{author}{Boardman, J.P.}, \bibinfo{author}{Rutherford, M.A.},
  \bibinfo{author}{Edwards, A.D.}, \bibinfo{author}{Hajnal, J.V.},
  \bibinfo{author}{Rueckert, D.}, \bibinfo{year}{2012}.
\newblock \bibinfo{title}{Construction of a consistent high-definition
  spatio-temporal atlas of the developing brain using adaptive kernel
  regression}.
\newblock \bibinfo{journal}{NeuroImage} \bibinfo{volume}{59},
  \bibinfo{pages}{2255--2265}.
\bibitem[{Sharma et~al.(2010)Sharma, Noblet, Rousseau, Heitz, Rumbach and
  Armspach}]{sharma2010evaluation}
\bibinfo{author}{Sharma, S.}, \bibinfo{author}{Noblet, V.},
  \bibinfo{author}{Rousseau, F.}, \bibinfo{author}{Heitz, F.},
  \bibinfo{author}{Rumbach, L.}, \bibinfo{author}{Armspach, J.P.},
  \bibinfo{year}{2010}.
\newblock \bibinfo{title}{Evaluation of brain atrophy estimation algorithms
  using simulated ground-truth data}.
\newblock \bibinfo{journal}{Medical image analysis} \bibinfo{volume}{14},
  \bibinfo{pages}{373--389}.
\bibitem[{Simonyan and Zisserman(2015)}]{simonyan2014very}
\bibinfo{author}{Simonyan, K.}, \bibinfo{author}{Zisserman, A.},
  \bibinfo{year}{2015}.
\newblock \bibinfo{title}{Very deep convolutional networks for large-scale
  image recognition}, in: \bibinfo{booktitle}{ICLR}.
\bibitem[{Sivera et~al.(2019)Sivera, Delingette, Lorenzi, Pennec, Ayache,
  Initiative et~al.}]{sivera2019model}
\bibinfo{author}{Sivera, R.}, \bibinfo{author}{Delingette, H.},
  \bibinfo{author}{Lorenzi, M.}, \bibinfo{author}{Pennec, X.},
  \bibinfo{author}{Ayache, N.}, \bibinfo{author}{Initiative, A.D.N.}, et~al.,
  \bibinfo{year}{2019}.
\newblock \bibinfo{title}{A model of brain morphological changes related to
  aging and alzheimer's disease from cross-sectional assessments}.
\newblock \bibinfo{journal}{NeuroImage} \bibinfo{volume}{198},
  \bibinfo{pages}{255--270}.
\bibitem[{Sullivan et~al.(1995)Sullivan, Marsh, Mathalon, Lim and
  Pfefferbaum}]{sullivan1995age}
\bibinfo{author}{Sullivan, E.V.}, \bibinfo{author}{Marsh, L.},
  \bibinfo{author}{Mathalon, D.H.}, \bibinfo{author}{Lim, K.O.},
  \bibinfo{author}{Pfefferbaum, A.}, \bibinfo{year}{1995}.
\newblock \bibinfo{title}{Age-related decline in mri volumes of temporal lobe
  gray matter but not hippocampus}.
\newblock \bibinfo{journal}{Neurobiology of aging} \bibinfo{volume}{16},
  \bibinfo{pages}{591--606}.
\bibitem[{Taubert et~al.(2010)Taubert, Draganski, Anwander, M{\"u}ller,
  Horstmann, Villringer and Ragert}]{taubert2010dynamic}
\bibinfo{author}{Taubert, M.}, \bibinfo{author}{Draganski, B.},
  \bibinfo{author}{Anwander, A.}, \bibinfo{author}{M{\"u}ller, K.},
  \bibinfo{author}{Horstmann, A.}, \bibinfo{author}{Villringer, A.},
  \bibinfo{author}{Ragert, P.}, \bibinfo{year}{2010}.
\newblock \bibinfo{title}{Dynamic properties of human brain structure:
  learning-related changes in cortical areas and associated fiber connections}.
\newblock \bibinfo{journal}{Journal of Neuroscience} \bibinfo{volume}{30},
  \bibinfo{pages}{11670--11677}.
\bibitem[{Taylor et~al.(2017)Taylor, Williams, Cusack, Auer, Shafto, Dixon,
  Tyler and Henson}]{camcan2015ageing}
\bibinfo{author}{Taylor, J.R.}, \bibinfo{author}{Williams, N.},
  \bibinfo{author}{Cusack, R.}, \bibinfo{author}{Auer, T.},
  \bibinfo{author}{Shafto, M.A.}, \bibinfo{author}{Dixon, M.},
  \bibinfo{author}{Tyler, L.K.}, \bibinfo{author}{Henson, R.N.},
  \bibinfo{year}{2017}.
\newblock \bibinfo{title}{{The Cambridge Centre for Ageing and Neuroscience
  ({Cam-CAN}) data repository: Structural and functional {MRI}, MEG, and
  cognitive data from a cross-sectional adult lifespan sample}}.
\newblock \bibinfo{journal}{NeuroImage} \bibinfo{volume}{144},
  \bibinfo{pages}{262--9}.
\bibitem[{Wang et~al.(2003)Wang, Simoncelli and Bovik}]{wang2003multiscale}
\bibinfo{author}{Wang, Z.}, \bibinfo{author}{Simoncelli, E.P.},
  \bibinfo{author}{Bovik, A.C.}, \bibinfo{year}{2003}.
\newblock \bibinfo{title}{Multiscale structural similarity for image quality
  assessment}, in: \bibinfo{booktitle}{The Thrity-Seventh Asilomar Conference
  on Signals, Systems \& Computers, 2003}, \bibinfo{organization}{{IEEE}}. pp.
  \bibinfo{pages}{1398--1402}.
\bibitem[{Wardlaw et~al.(2015)Wardlaw, Vald{\'e}s~Hern{\'a}ndez and
  Mu{\~n}oz-Maniega}]{wardlaw2015white}
\bibinfo{author}{Wardlaw, J.M.}, \bibinfo{author}{Vald{\'e}s~Hern{\'a}ndez,
  M.C.}, \bibinfo{author}{Mu{\~n}oz-Maniega, S.}, \bibinfo{year}{2015}.
\newblock \bibinfo{title}{What are white matter hyperintensities made of?
  relevance to vascular cognitive impairment}.
\newblock \bibinfo{journal}{Journal of the American Heart Association}
  \bibinfo{volume}{4}, \bibinfo{pages}{e001140}.
\bibitem[{Wegmayr et~al.(2019)Wegmayr, H{\"o}rold and
  Buhmann}]{wegmayr2019generative}
\bibinfo{author}{Wegmayr, V.}, \bibinfo{author}{H{\"o}rold, M.},
  \bibinfo{author}{Buhmann, J.M.}, \bibinfo{year}{2019}.
\newblock \bibinfo{title}{{Generative Aging of Brain MR-Images and Prediction
  of Alzheimer Progression}}, in: \bibinfo{booktitle}{German Conference on
  Pattern Recognition}, \bibinfo{organization}{Springer}. pp.
  \bibinfo{pages}{247--260}.
\bibitem[{Westlye et~al.(2009)Westlye, Walhovd, Dale, Espeseth, Reinvang, Raz,
  Agartz, Greve, Fischl and Fjell}]{westlye2009increased}
\bibinfo{author}{Westlye, L.T.}, \bibinfo{author}{Walhovd, K.B.},
  \bibinfo{author}{Dale, A.M.}, \bibinfo{author}{Espeseth, T.},
  \bibinfo{author}{Reinvang, I.}, \bibinfo{author}{Raz, N.},
  \bibinfo{author}{Agartz, I.}, \bibinfo{author}{Greve, D.N.},
  \bibinfo{author}{Fischl, B.}, \bibinfo{author}{Fjell, A.M.},
  \bibinfo{year}{2009}.
\newblock \bibinfo{title}{Increased sensitivity to effects of normal aging and
  alzheimer's disease on cortical thickness by adjustment for local variability
  in gray/white contrast: a multi-sample mri study}.
\newblock \bibinfo{journal}{Neuroimage} \bibinfo{volume}{47},
  \bibinfo{pages}{1545--1557}.
\bibitem[{Woolrich et~al.(2009)Woolrich, Jbabdi, Patenaude, Chappell, Makni,
  Behrens, Beckmann, Jenkinson and Smith}]{woolrich2009bayesian}
\bibinfo{author}{Woolrich, M.W.}, \bibinfo{author}{Jbabdi, S.},
  \bibinfo{author}{Patenaude, B.}, \bibinfo{author}{Chappell, M.},
  \bibinfo{author}{Makni, S.}, \bibinfo{author}{Behrens, T.},
  \bibinfo{author}{Beckmann, C.}, \bibinfo{author}{Jenkinson, M.},
  \bibinfo{author}{Smith, S.M.}, \bibinfo{year}{2009}.
\newblock \bibinfo{title}{Bayesian analysis of neuroimaging data in {FSL}}.
\newblock \bibinfo{journal}{Neuroimage} \bibinfo{volume}{45},
  \bibinfo{pages}{S173--S186}.
\bibitem[{Xia et~al.(2019)Xia, Chartsias, Tsaftaris and
  Initiative}]{xia2019consistent}
\bibinfo{author}{Xia, T.}, \bibinfo{author}{Chartsias, A.},
  \bibinfo{author}{Tsaftaris, S.A.}, \bibinfo{author}{Initiative, A.D.N.},
  \bibinfo{year}{2019}.
\newblock \bibinfo{title}{Consistent brain ageing synthesis}, in:
  \bibinfo{booktitle}{International Conference on Medical Image Computing and
  Computer-Assisted Intervention}, \bibinfo{organization}{Springer}. pp.
  \bibinfo{pages}{750--758}.
\bibitem[{Zecca et~al.(2004)Zecca, Youdim, Riederer, Connor and
  Crichton}]{zecca2004iron}
\bibinfo{author}{Zecca, L.}, \bibinfo{author}{Youdim, M.B.},
  \bibinfo{author}{Riederer, P.}, \bibinfo{author}{Connor, J.R.},
  \bibinfo{author}{Crichton, R.R.}, \bibinfo{year}{2004}.
\newblock \bibinfo{title}{Iron, brain ageing and neurodegenerative disorders}.
\newblock \bibinfo{journal}{Nature Reviews Neuroscience} \bibinfo{volume}{5},
  \bibinfo{pages}{863}.
\bibitem[{Zhang et~al.(2001)Zhang, Brady and Smith}]{zhang2001segmentation}
\bibinfo{author}{Zhang, Y.}, \bibinfo{author}{Brady, M.},
  \bibinfo{author}{Smith, S.}, \bibinfo{year}{2001}.
\newblock \bibinfo{title}{Segmentation of brain mr images through a hidden
  markov random field model and the expectation-maximization algorithm}.
\newblock \bibinfo{journal}{TMI} \bibinfo{volume}{20}, \bibinfo{pages}{45--57}.
\bibitem[{Zhang et~al.(2016)Zhang, Shi, Wu, Wang, Yap and
  Shen}]{zhang2016consistent}
\bibinfo{author}{Zhang, Y.}, \bibinfo{author}{Shi, F.}, \bibinfo{author}{Wu,
  G.}, \bibinfo{author}{Wang, L.}, \bibinfo{author}{Yap, P.T.},
  \bibinfo{author}{Shen, D.}, \bibinfo{year}{2016}.
\newblock \bibinfo{title}{Consistent spatial-temporal longitudinal atlas
  construction for developing infant brains}.
\newblock \bibinfo{journal}{TMI} \bibinfo{volume}{35},
  \bibinfo{pages}{2568--2577}.
\bibitem[{Zhang et~al.(2017)Zhang, Song and Qi}]{zhang2017age}
\bibinfo{author}{Zhang, Z.}, \bibinfo{author}{Song, Y.}, \bibinfo{author}{Qi,
  H.}, \bibinfo{year}{2017}.
\newblock \bibinfo{title}{Age progression/regression by conditional adversarial
  autoencoder}, in: \bibinfo{booktitle}{{CVPR}}, pp.
  \bibinfo{pages}{5810--5818}.
\bibitem[{Zhao et~al.(2019)Zhao, Adeli, Honnorat, Leng and
  Pohl}]{zhao2019variational}
\bibinfo{author}{Zhao, Q.}, \bibinfo{author}{Adeli, E.},
  \bibinfo{author}{Honnorat, N.}, \bibinfo{author}{Leng, T.},
  \bibinfo{author}{Pohl, K.M.}, \bibinfo{year}{2019}.
\newblock \bibinfo{title}{Variational autoencoder for regression: Application
  to brain aging analysis}.
\newblock \bibinfo{journal}{MICCAI} .
\bibitem[{Zhu et~al.(2017)Zhu, Park, Isola and Efros}]{zhu2017unpaired}
\bibinfo{author}{Zhu, J.Y.}, \bibinfo{author}{Park, T.},
  \bibinfo{author}{Isola, P.}, \bibinfo{author}{Efros, A.A.},
  \bibinfo{year}{2017}.
\newblock \bibinfo{title}{Unpaired image-to-image translation using
  cycle-consistent adversarial networks}, in: \bibinfo{booktitle}{ICCV}, pp.
  \bibinfo{pages}{2223--2232}.
\bibitem[{Ziegler et~al.(2012)Ziegler, Dahnke and Gaser}]{ziegler2012models}
\bibinfo{author}{Ziegler, G.}, \bibinfo{author}{Dahnke, R.},
  \bibinfo{author}{Gaser, C.}, \bibinfo{year}{2012}.
\newblock \bibinfo{title}{Models of the aging brain structure and individual
  decline}.
\newblock \bibinfo{journal}{{Frontiers in Neuroinformatics}}
  \bibinfo{volume}{6}, \bibinfo{pages}{3}.

\end{thebibliography}

\end{document}